# Equilibrium nano-shape change induced by epitaxial stress
## (Generalised Wulf-Kaishew theorem)

P.Müller, R.Kern

*Centre de Recherche sur les Mécanismes de la Croissance Cristalline[1], CRMC2-CNRS, Campus de Luminy, case 913, F-13288 Marseille Cedex 9, France*

## Abstract

A generalised Wulf-Kaishew theorem is given describing the equilibrium shape (ES) of an isolated 3D crystal A deposited coherently onto a lattice mismatched planar substrate. For this purpose a free polyhedral crystal is formed then homogeneously strained to be accommodated onto the lattice mismatched substrate. During its elastic inhomogeneous relaxation the epitaxial contact remains coherent so that the 3D crystal drags the atoms of the contact area and produces a strain field in the substrate. The ES of the deposit is obtained by minimising at constant volume the total energy (bulk and surface energies) taking into account the bulk elastic relaxation. Our main results are: **(1)** Epitaxial strain acts against wetting (adhesion) so that globally it leads to a thickening of the ES. **(2)** Owing to strain the ES changes with size. More precisely the various facets extension changes, some facets decreasing, some others increasing. **(3)** Each dislocation entrance, necessary for relaxing plastically too large crystals abruptly modifies the ES and thus the different facets extension in a jerky way. **(4)** In all cases the usual self-similarity with size is lost when misfit is considered. We illustrate these points in case of box shaped and truncated pyramidal crystals. Some experimental evidences are discussed.

---

[1] Associé aux Universités Aix-Marseille II et III.

**I/ Introduction**:

Macroscopic crystal shape studies founded crystal physic. Genuine crystal growers consider them however as an academic game in spite of the fact they do not ignore that such studies reveal essential growth mechanisms they need [1-4]. Thin film epitaxial growth, also based on geometry [5], quickly took great advantage of *shape studies but on nanoscale* whose impact became vital for high integration circuitry. The different epitaxial growth modes (shapes) [6,7] influence defect entrance, segregation [8] bringing with them either deleterious or benefic physical effects depending what applications are set as a goal. Coupled morphology-growth mechanism studies combined with in-situ surface physics techniques developed in the last decade with intense technological activities on Si, Ge or III-V semi-conductors. From these resulted several scientific discoveries as the effect of strain on surface morphology [9-13]. Such by technology stimulated studies may they be qualified as too academic ?

In this paper therefore we deliberately revisited the 100 years-old academic Wulf-theorem [14,22] concerning the equilibrium shape (ES) of free crystals or more exactly the 50 years-old Kaishew's theorem [15] saying how a substrate influences Wulf's shape. *Our topic is to introduce epitaxial strain as an ingredient in the classic corpus of ES crystals*. Indeed Kaishew's theorem [15] does not consider any lattice mismatch in between the substrate and its deposit. Thus it only describes correctly the ES of deposited crystal in case of non coherent epitaxy (as glissile or Van der Waals epitaxy [17-21]) or in case of coherent epitaxy but with zero misfit. *For 3D coherent epitaxies* on a lattice-mismatched substrate the deposited crystal is strained as well as a part of the underlying substrate [10,11,29,31] . Thus since the mechanical equilibrium of the supported crystal is reached when its free surfaces have vanishing normal stress components [23], the elastic energy density changes with the shape of the crystal and thus can be minimal for a specific shape at a given volume. In other words the ES must depend on epitaxial strain, as it has been foreseen by some theoretical works for deposited solid drops [26,27], pyramids [28,29], box shaped crystals [24] but under some restrictive conditions or models. A recent very general theoretical paper [30] concludes even very generally: " the shape of the strained particle may bear little resemblance to more classical Wulf shape". In this paper we want to analyse in a general but comprehensive way the ES changes induced by elasticity. For this purpose in II we give a *generalised Wulf-Kaishew theorem* describing the polyhedral ES of epitaxially strained crystals. More precisely we show that according to the epitaxial strain value and crystal size, some facets can appear or disappear so that self-similarity of the usual ES is no more preserved. Then for illustration in III we apply the theorem and construct the ES for two specific cases: box shaped crystal and truncated pyramid. We describe quantitatively the shape changes with size, misfit, adhesion to substrate and the relative substrate to deposit stiffness. In II3 and for each case in III3 we also consider the ES change induced by dislocation entrance as we preliminary reported in [25]. At last in IV we discuss (IV1) some weak points of former works and compare our results with experimental evidences (IV2).

**II/ Towards a generalised Wulf-Kaishew theorem**

    **II.1/ Thermodynamical process**

The ES of epitaxially strained crystal is found by minimising the total free energy $\Delta F$ needed to form a 3D crystal A onto a lattice-mismatched substrate B. For this purpose, the thermodynamical process depicted on figure 1 is useful. In a first time a polyhedral crystal is formed from an infinite reservoir of crystalline matter A. In a second time this crystal A is homogeneously strained to be accommodated on its basis face on the stress free substrate B. This elastic state is not a minimum state of energy. All stress components normal to the surfaces have to vanish so the system must relax (third time). During this elastic relaxation the deposited crystal A, supposed to remain coherent to its substrate, drags the atoms of the contact area and produces a strain field in its underlying substrate B. So even if the total elastic energy is lowered by relaxation the elastic energy density in the 3D crystal has effectively been lowered whereas in the substrate it increased. After elastic relaxation the 3D crystal and its substrate are inhomogeneously strained. Finally the crystal shape of A has to be changed at constant number of atoms, the self-consistent interplay of surface change and elastic relaxation leading to the ES. The total free energy change of the thermodynamical process of figure 1 can thus be written as the sum of three terms.

* The first term is the *chemical work* spent to form the crystal A (volume V) from the infinite reservoir of A. It reads

$$\Delta F_1 = -\Delta\mu V \qquad (1)$$

where $\Delta\mu$ is called supersaturation per unit volume. For a perfect vapour A at the pressure P in respect to the saturation pressure $P_\infty$ of the infinite reservoir, it reads $\Delta\mu = \frac{kT}{v}\ln\left(\frac{P}{P_\infty}\right)$, v being the volume of a molecule in A.

* The second term corresponds to the *formation of surfaces and interfaces*. For a crystal having i facets of area $S_i$ characterised by their surface energies $\gamma_i S_i$ (see figure 2a) this term reads

$$\Delta F_2 = \sum_i \gamma_i S_i + S_{AB}(\gamma_{AB} - \gamma_B) \qquad (2)$$

where the summation is carried out on the *free surfaces* of A. $\gamma_{AB}$ is the interfacial energy density, $S_{AB}$ the contact area and $\gamma_B$ the surface energy density of the free face of the substrate B having been exchanged by AB.

* The third term is the *elastic energy stored by the relaxed system* (partially relaxed deposit + strained substrate). For a biaxially strained crystal the elastic energy *before relaxation* is $E_o m^2 V$ where $m = (b-a)/a$ is the epitaxial misfit in between A (parameter a) and B (parameter b), $E_o$ is a combination of elastic coefficients of A and V the volume

of the deposited crystal A. Owing to the fact that the relaxation lowers the elastic energy, the elastic energy finally stored by the relaxed system reads:

$$\Delta F_3 = \mathbf{E}_o m^2 V R \qquad (3)$$

where $0 < R < 1$ is a relaxation energy factor that *must depend in a complex way upon the crystal shape*. Nevertheless its limiting behaviour must be $R = 0$ for a completely relaxed system and $R = 1$ for a non relaxed crystal. More precisely since the deposit is coupled to the substrate, during relaxation it stresses the underlying crystal so that the relaxation factor $R$ contains two contributions. The first is due to the deposit A and reads $R_A$, the second is due to the substrate B and reads $R_B$ so that there is

$$R = R_A + R_B \qquad (4)$$

According to (3,4) the elastic energy localises one part in the deposit and the other part in the substrate; both being proportional to misfit square and to volume V. This is an essential characteristic of coherent epitaxies. Furthermore the relaxation factors $R_A$ and $R_B$ must also depend upon the relative rigidity K of the substrate in respect to the deposit. For an infinitely rigid substrate the deposit relaxes ($R_A \neq 0$) but does not drag its substrate so that there is $R_B = 0$. On the contrary for infinitely weak substrate the deposited crystal may completely relax ($R_A = 0$) but the underlying substrate does not store any elastic energy so that there is also $R_B = 0$. In fact the relaxation factors $R_A$, $R_B$ and thus $R$, have to be calculated for each shape, relative rigidity and crystalline orientation as well. This has been done for a box shaped crystal in [24]. We will come back in detail to specific R factors in III1.

The total free energy change induced by the thermodynamical process depicted in figure 1 thus reads:

$$\Delta F = \Delta F_1 + \Delta F_2 + \Delta F_3 = -\Delta \mu V + \sum_{i \neq AB} \gamma_i S_i + S_{AB}(\gamma_{A-} \beta) + \mathbf{E}_o m^2 V R \qquad (5)$$

where we have used Dupré's relation $\gamma_{AB} = \gamma_A + \gamma_B - \beta$ where $\beta$ is the adhesion energy, and $\gamma_A$ the surface energy of the basal face of A before adhesion onto B.

Lastly let us stress on the fact that in the former process we have neglected surface energy changes induced by the epitaxial strain so that the elastic term in (5) is a pure bulk property. In fact two main effects of strain on the surface energy can be encountered. The first one is a continuous change of surface energy with strain due to surface stress. Such effect can be taken into account by adding to (5) a surface elasticity term (surface stress work against bulk deformation of the various surfaces and interfaces) scaling linearly with strain as we have done in a complementary paper [63]. Nevertheless since (i) the strain-induced gamma-plot change remains weak[2] and (ii) the bulk stress relaxation takes place by the free surfaces of the crystal, the surface stress effect on the ES remains weak as yet depicted by [47] and discussed in [63]. The second effect of strain on surface energy may be due to discontinuous change of surface energy in some crystallographic orientation with stress due to strain-stabilised surface reconstructions. Such surface transitions may severely alter the surface energy and thus modify the ES as we will discuss separately in section IV22.

**II2/ Shape change induced by coherent epitaxial strain**

The ES is reached when the first order differential $\Delta F$ of (5)

$$d\Delta F = -\Delta\mu dV + \sum_{i \neq AB} \gamma_i dS_i + (\gamma_A - \beta)dS_{AB} + \mathbf{E}_o m^2 V dR + \mathbf{E}_o m^2 R dV \quad (6)$$

vanishes at constant volume.

For calculating (6) some points have to be underlined:

(i) The volume V of a free polyhedral crystal can be split into pyramids of heights $h_i$

---

[2] for instance for InAs dots grown on GaAs(001), at equilibrium surface stress lowers the surface energy by less than 10% for {101} {111} and {-1-1-1} faces according to [47].

and basis $S_i$ so that $V = \frac{1}{3}\sum_i h_i S_i$ and thus up to the first order $dV = \frac{1}{2}\sum_i h_i dS_i$. For an adhering crystal the summation has to be restricted to the free faces i≠AB so that the emerging volume writes for a variation $dV = \frac{1}{2}\sum_{i \neq AB} h_i dS_i + \frac{1}{2} h_{AB} dS_{AB}$. The distance $h_{AB}$ is that of the common pyramid summits (Wulf point) to the interface AB, being taken negative if the Wulf point is inside the substrate (as in figure 2a), and positive if located outside.

(ii) Since the crystal is a convex surface there is $S_A = S_{AB} - \sum_{i \neq A,AB} S_i \cos\theta_i$ where $S_A$ is the surface area of the top face of the crystal, $\theta_i$ the angle of the other faces in respect to the substrate (see figure 2 where i=1,2 with $\theta_i = \theta_1 \neq \pi/2$ and $\theta_2 = \pi/2$ ).

(iii) The relaxation factor $R$ depends on the shape of the crystal and thus must depend on the surface areas $S_i$ and $S_{AB}$.

With all these considerations dΔF of (6) can be written:

$$d\Delta F = -\frac{1}{2}\left(\Delta\mu - \mathbf{E}_o m^2 R\right)\left[\sum_{i \neq A,AB} h_i dS_i + h_A\left(dS_{AB} - \sum_{i \neq A,AB} \cos\theta_i dS_i\right)\right] +$$
$$\sum_{i \neq A,AB} \gamma_i dS_i + (\gamma_A - \beta)dS_{AB} + \gamma_A\left(dS_{AB} - \sum_{i \neq A,AB} \cos\theta_i dS_i\right) + h_{AB} dS_{AB} +$$
$$\mathbf{E}_o V m^2 \left[\sum_{i \neq A,AB} \left.\frac{\partial R}{\partial S_i}\right|_{S_{AB}} dS_i + \left.\frac{\partial R}{\partial S_{AB}}\right|_{S_i} dS_{AB}\right] \quad (7)$$

The ES thus is obtained when all the partial derivatives become zero simultaneously, that means $\left.\frac{\partial \Delta F}{\partial S_i}\right|_{AB} = 0$ (i≠A,AB) and $\left.\frac{\partial \Delta F}{\partial S_{AB}}\right|_i = 0$. Collecting them from (7) leads to the set of i connected equations describing the free faces i:

$$\frac{\Delta\mu - \mathbf{E}_o m^2 R}{2} == \frac{\gamma_i - \gamma_A \cos\theta_i + \mathbf{E}_o m^2 \frac{V}{n_i}\left.\frac{\partial R}{\partial S_i}\right|_{AB}}{h_i - h_A \cos\theta_i} \quad (8)$$

and the following one describing the substrate truncation

$$\frac{\Delta\mu - \mathbf{E}_o m^2 R}{2} = \frac{2\gamma_A - \beta + \mathbf{E}_o m^2 V \left.\frac{\partial R}{\partial S_{AB}}\right|_i}{H} \quad (8')$$

valid for all i≠A,AB.

In equation (8) $n_i$ is the number of crystallographic equivalent faces i of area $S_i$ on the crystal and in (8') $H = h_A + h_{AB}$ is the total crystal height above the substrate (see figure 2a). We call this set of equations (8), (8') generalised Wulf-Kaishew theorem

Let us discuss in details at (i), (ii) and (iii) the implications of this theorem

*(i)Wulf-Kaishew theorem (m=0)*

In absence of misfit, m=0, relation (8) is now valid for all faces and reads

$$\lambda = \frac{\Delta\mu}{2} = \frac{\gamma_i}{h_i}, \qquad i \neq AB \quad (9)$$

It is nothing else than Wulf's theorem [14,22] valid for all singular faces i of a free non-supported crystal A and corresponding to the i inward cusps of its gamma plot. Taking for the purpose λ=1, $h_i = \gamma_i$, we build around a point O (Wulf's construction) the minimal polyhedron. *For different sizes $\lambda$, the ES are self similar.* (See figure 3a for illustration). Relation (8') gives a supplementary relation to (9)

$$\lambda = \frac{\Delta\mu}{2} = \frac{2\gamma_A - \beta}{H} \quad (9')$$

called Kaishew's theorem [15,32,33]. It concerns the substrate and says that for increasing adhesion energy β of the basal face A on the flat substrate B, the ES is preserved since (9) is still valid, but the crystal A is thus more « truncated by the substrate » β approaches the value $2\gamma_A$. In other words *for a given volume of A* there is an « emerging height » $H = 2\gamma_A - \beta$, becoming for β=0, $H = 2h_A$ that means the full height of the free non supported ES crystal. When $\beta \rightarrow 2\gamma_A$ the emerging height vanishes. Half truncation $H = h_A$ happens when $\beta = \gamma_A$, so that one half of the crystal A is real the other half part virtually immersed in the substrate.

*For different emerging volumes* there is *again self-similarity* but the similarity centre is a point S located at the interface. See figures 3 b,c,d. The different volumes are no more concentric since from (9) (9')

$$r_i = \frac{H}{h_i} = \frac{2\gamma_A - \gamma_i}{\gamma_i} \quad \text{for all i} \neq \text{AB} \tag{10}$$

The Wulf point O is located at a distance OS=$h_{AB}$ from the interface:

$$h_{AB} = \left(1 - \frac{\beta}{\gamma_A}\right) h_A \tag{11}$$

Their volumes are self similar so that for $\lambda \neq 1$

$$h_i \to \lambda h_i \quad \text{for all i} \neq \text{AB} \tag{12}$$

but Wulf's point O wanders from OS=$h_{AB}$ to O'S=$h'_{AB}$ since from (12) (11) there is also

$$h_{AB} \to \lambda h_{AB} \tag{13}$$

So the Wulf point wanders either outside or inside the substrate according to $\beta < \gamma_A$ or $\beta > \gamma_A$ respectively. There is no common Wulf point for crystals of various sizes but self similarity from point S. Only when $\beta = \gamma_A$ the Wulf points O,O', O'' of all these different volumes coalesce in one unique point S (see figure 3c) which becomes the similarity centre of all volumes. When $\beta \neq \gamma_A$ this point S preserves latter property what is illustrated by the « growth sectors » issued from S, but this needs a proof we give in appendix A.

(ii) When there is a lattice mismatch (m≠0) but *when elastic relaxation is neglected* the relaxation factor is $R = 1$ and all its partial derivatives therefore vanish in (8, 8'). The ES remains self similar since (10) is still valid. Nevertheless in this case three dimensional (3D) growth only is possible for $h_i, h_{AB} > 0$ that means when the supersaturation per atom $\Delta\mu$ overpasses the bulk energy density $\mathbf{E}_o m^2$ of the fully strained crystal [24,34]

(iii) *Generalised Wulf Kaishew theorem*

When m≠0 and when elastic relaxation is taken into account, the aspect ratios characterising the ES become from (8,8'):

$$r_i = \frac{H}{h_i - h_A \cos\theta_i} = \frac{2\gamma_A - \beta + \mathbf{E}_o m^2 V \left.\frac{\partial R}{\partial S_{AB}}\right|_{S_i}}{\gamma_i - \gamma_A \cos\theta_i + \mathbf{E}_o m^2 \frac{V}{n_i}\left.\frac{\partial R}{\partial S_i}\right|_{S_{AB}}} \tag{14}$$

where $H = h_A + h_{AB}$ again is the crystal height and $\Lambda = h_i - h_A \cos\theta_i$ a measure of the top face A limited by the i face (see figure 2a).

Clearly the partial derivatives of the relaxation factor R which now appear in (14) change the ES. More precisely the partial derivative $\left.\frac{\partial R}{\partial S_{AB}}\right|_{S_i \neq A}$ describes the elastic energy density change versus interfacial area change. Since an extension of the interfacial area must increase the elastic energy (let us recall that for an infinite coherent uniform film no relaxation occurs) there must be $\left.\frac{\partial R}{\partial S_{AB}}\right|_{S_i \neq A} > 0$ (see figure 2b). On the contrary, for a given interface area, each facet extension helps to elastic relaxation (let us recall that 3D epitaxially strained crystals relax by their free edges) so that $\left.\frac{\partial R}{\partial S_i}\right|_{S_{AB}} < 0$ (see figure 2c). *Thus the aspect ratios (14) that characterise the crystal shape increase as a function of the volume and thus there is no more similarity of the ES*. More precisely for a given volume the epitaxial strain thickens the ES (the total height H increases and the lateral size decreases). Furthermore since the different derivatives $\left.\frac{\partial R}{\partial S_i}\right|_{S_{AB}} < 0$ of various non crystallographic equivalent faces i have no physical reason to be the same the different central distances $h_i$ do not behave similarly with $E_o m^2 V$ (see (14)) that means with misfit strain m and volume V. *So the ES must change continuously with volume*. More precisely the relative facet extension must change during near equilibrium growth, some facets decreasing, some others increasing as it will be illustrated quantitatively in the section III. Nevertheless let us stress again that since stress-induced change of the surface energy have not been considered, theorem (8) only gives access to strain-induced ES changes *at constant gamma-plot*. In other words facets of new crystallographic orientation cannot be created by bulk elasticity. We will see in section IV22 that it may be different when strain induced change of surface energy are considered.

**II3/ Shape change induced by dislocation entrance**

We have just seen that the ES continuously changes when increasing its size. However the system accumulates strain energy that may become prohibitive so that plastic relaxation proceeds. We consider here the case of dislocation entrance.

Each dislocation entrance leads to an abrupt change in strain. This drop of strain can roughly be obtained from geometrical (vernier) considerations as classicaly used by Jesser [64] and Matthews [8] so that when there are N interfacial dislocations the released misfit in the directions i=x,y of the interfacial plane reads

$$m_i^{'} = m_i - N \frac{b}{l_N^i}$$

where $l_N^i$ is the interfacial width of the crystal having N interfacial dislocations with Burger vector component b parallel to direction i. In the following and for the sake of simplicity we only consider quadratic crystals sit on their basis containing a double array of identical dislocations which may appear thermodynamically simultaneously[3]. Thus one can define the

---

[3] Obviously form a kinetics point of view the dislocations may enter latter on and not simultaneously.

single misfit

$$m' = m - N\frac{b}{\sqrt{S_{AB}}\big|_N} \quad (15)$$

where $S_{AB}\big|_N$ is the interfacial area for a crystal having N interfacial dislocations. Though rough expression (15) takes account the discrete nature of the dislocation-induced strain relaxation. Furthermore as reported by Matthews [8] expression (15) reproduces the discrete change of strain observed by R.Vincent on β Sn(010)/SnTe(001)[65].

Theorem (8) shows that elastic energy modifies the ES. Thus since each dislocation entrance leads to a discrete drop of strain, each dislocation entrance may abruptly modify the ES. In order to illustrate this point we will proceed in a schematic way to define a thermodynamical criterion giving the condition where dislocations are allowed to enter. Then we discuss the effect of the dislocation entrance on the ES. For this purpose we deliberately ignore the details and refinement of the various models of relaxation by dislocation entrance and only emphasise on the minimal basic ingredients leading to reasonable results. More refined models including elastic interaction in between dislocations, dislocation images background, elastic fields resulting of dislocation formation etc… exist but do not modify the main results about the effect of dislocation entrance on the ES.

From a thermodynamical point of view the number of interfacial dislocations may pass from N to N+1 when the total elastic bulk energy change due to the introduction of the (N+1)[th] dislocation is negative (again we neglect surface stress effects). The elastic energy stored by the system when N interfacial dislocations exist can shorthand be written following the basic ideas of Matthews [8]

$$W_N = 2\frac{m-m'}{a}b^2\frac{E_{AB}}{2\pi}(1+\ln\chi)S_{AB} + \mathsf{E}_o m'^2 VR \quad (16)$$

where the first term is the self energy of a crossed array of perpendicular non interacting interfacial dislocations with $1/E_{AB} = 1/E_A + 1/E_B$ some reciprocal interfacial modulus[4] ($E_A$ and $E_B$ are the elastic moduli of A and B respectively), b the Burger's vector modulus component in the interface, m-m'>0 the part of the misfit accommodated by the (N+1) dislocations and χ a cut-off radius around the dislocation. When H<d where d is the equidistance in between dislocations there is χ=H/2a, if not there is χ=d/2a [8]. The second term in (16) is the elastic energy stored in 3D island relaxed elastically and plastically by N dislocations having reduced the misfit from m to m' rouhghly given by (15).

The thermodynamical criterion for the (N+1)$^{th}$ dislocation entrance is thus obtained from $W_{N+1} - W_N < 0$ that means with (15) and (16) when the critical height fulfils:

$$H_c = \frac{V}{S_{AB}|_N} > \frac{b^2}{a} \frac{1}{2\pi} \frac{K}{1+K} \frac{1+\ln(\chi)}{\left[m - \frac{N+1/2}{\sqrt{S_{AB}|_N}/b}\right]R} \quad (17)$$

where $K = E_B/E_A$ is the relative rigidity of the substrate B in respect to the deposited crystal. Let us discuss (17)

(i) For a pseudomorphous uniform film $\sqrt{S_{AB}|_N} \to \infty$, R=1 and $V/S_{AB}|_N = H_c^\infty$ is the usual critical film thickness beyond which a dislocation may enter as find by Mathews [8].

(ii) For a 3D crystal ($\sqrt{S_{AB}|_N} \neq \infty, R<1$) there are different values of the interfacial area beyond which 1, 2...j dislocations may introduce. More precisely for an increasing crystal there exists a ratio $V/S_{AB}|_1$ where the first dislocation may enter. This first dislocation entrance lowers the misfit from m to $m - b/\sqrt{S_{AB}|_1}$ (see 16). However since we have seen in the previous section that the ES of a crystal is misfit dependent the misfit reduction due to dislocation entrance must change the ES of the crystal. More precisely *each dislocation*

---

[4] In fact a so-defined modulus $E_{AB}$ is fictious. Its introduction is a simplified way to tell that dislocations preferentially stay in the softest part of the bi-crystal.

*entrance lowers the misfit m and thus lowers the aspect ratios (14) leading to a flattening of the ES of a deposited crystal.*

At this point we can summarise our main general results about ES change due to misfit and dislocation entrance.

*(1) The epitaxial stress acts against wetting (adhesion) so that globally it leads to a thickening of the ES*

*(2) Owing to the coherent strain the ES changes with size so that the ES continuously changes during near equilibrium epitaxial growth. There is no more self-similar as in the strain free deposit (m=0) case.*

*(3) Each dislocation entrance necessary for relaxing too large crystals abruptly modifies the ES.*

**III/ Application: construction of the ES**

Our generalised Wulf-Kaishew theorem (8), (8') can only be practically useful when (i) the gamma plot $\gamma_A(n)$ is known, (ii) the adhesion energy β is known and for avoiding complications there is no mixing, (iii) the relaxation factor R is known for various shapes. The procedure used to calculate the ES thus consists in injecting the relaxation factor R of (3) and then to solve the generalised Wulf-Kaishew theorem (8), (8') in terms of aspect ratios. In the following we will apply the theorem to quadratic crystals. For this purpose we choose two quadratic crystal shape families. We take a $\gamma_A(n)$ plot of the free crystal which contains a double set of eight Herring spheres having a common Wulf point O (See in figure 4a projection along the (100) plane containing the quaternary axis [001] and the binary axis [010]). Each subset has a leading sphere of diameter OA and OB making respectively an angle α/2 and β/2 with these axis. By symmetry the full sets are generated defining mutual intersections, especially 14 inward cusps. These cusps will lead for a free crystal to the morphology {001}2, {011}8, {010}4 that means 2 basal faces, a bipyramid of 8 faces and a prism of 4 faces[5]. These are the singular faces of this quadratic crystal originating from this gamma plot. Owing to adhesion on its (001) plane and in agreement with the usual Wulf Kaishew theorem the ES results from a truncation parallel to the (001) face at a level H from the top face. *In the following we will limit ourselves to the value H given in figure 4a so that it results a truncated pyramid.* When the two leading spheres degenerate, their diameters OA and OB becoming one, ($\alpha+\beta \rightarrow \pi$), it results the gamma plot of figure 4b having a simple set of eight spheres defining 6 inwards cusps. For a non-supported crystal they lead from Wulf's construction to a quadratic box {010}4 that means a quadratic prism and two basal faces {001}2. Owing to adhesion with (001) plane the substrate truncates the ES at height H.

The relaxation factors $R$ of these two emerging shapes (fig. 4a,4b) are available (see section III.1), so that the strained ES (see section III.2) can be quantitatively calculated. Thus in the two cases we can predict the effect of misfit, wetting and initial facet extension on the strained ES (see section III.2). At last their ES change induced by dislocation entrance (see section III.3) can also be predicted.

### III.1/ Relaxation factor

In figures 4c and 4d we plot the known relaxation factors of each of these shapes families when coherently accommodated onto a mismatched substrate. The unique variable is $r = H/l$ the shape ratio defined as the height over the basal length. For box shaped crystals (fig.4d) the relaxation factors $R_A(r), R_B(r)$ of the deposit A and substrate B respectively and thus $R(r) = R_A(r) + R_B(r)$ are calculated ones [31,24]. First, the equilibrium strain components in an epitaxially strained ribbon in the framework of *continuous isotropic elasticity* are calculated in a self-consistent way following Hu [38]. The elastic energy per unit length then is calculated bringing a relaxation factor depending on the aspect ratio $r = H/l$. By the superposition principle the energy of a relaxed box shaped crystal is calculated. Corners effects are of course not considered by this way. In the expression of these aspect ratios appears, as a parameter, the relative rigidity K of the substrate in respect to deposit. In figure 4d we plot these relaxation factors for K=1. It is seen that when r=0, a uniform coherent film is not relaxed ($R_A(0) = 1$) and therefore a primarily non-stressed substrate does not become stressed too ($R_B(0) = 0$). However as the film becomes a rectangular box, finite r, it relaxes from its borders so that $R_A(r) < 1$ and therefore it stresses the underlying substrate where now $R_B(r) \neq 0$. While $R_A(r)$ tends towards zero for very high and narrow boxes (since only the very basal parts are stressed) the substrate becomes less stressed and $R_B(r) \to 0$. Therefore while $R_A(r)$ decreases, $R_B(r)$ passes through a maximum. Nevertheless the total relaxation factor $R(r)$ continuously decreases. From figure 4d where K=1 one can argue that a flat box, i.e r=0.1, lost 70% of its initial strain energy by relaxation, or it stores only 30%, the substrate however storing 20%. For a box, of r=0.2, both deposit and substrate bear the same energy, higher boxes have an inversion of this partition. For instance, for r=0.5 there is 6% in the deposit and 10% in the substrate. From the formulae (4-6) of [31,24] one can easily calculate the effect of relative stiffness K of the substrate. For K<1, weaker substrate $R_A(r)$ becomes steeper and $R_B(r)$ higher. It is the contrary for K>1.

---
[5] For manipulation of gamma plots and the related ES refer to [22,35-37]

\* For truncated pyramids we have not been able to calculate by the same method the relaxation factors. Ponchet et al [39], among others (whose results are discussed in IV1.2), have used numerical methods to calculate the elastic energy of relaxed truncated pyramids A coherently strained on a lattice mismatched substrate B. Using finite elements methods they determined the distribution of elastic energy within the truncated square based pyramids as well as within the underlying coherent substrate. Was considered the specific case where

substrate and deposit approximately have the same cubic elastic properties (K≈1) and the pyramid an angle of θ=45°. They give the relaxation factor $R(r) = R_A(r) + R_B(r)$ for a wide and useful range of aspect ratios from severely truncated pyramids (r=0.044) to full pyramid ones (r=0.5). They showed that the results are not dependent upon the volume of the 3D crystal. Partial relaxation factors $R_A(r)$ and $R_B(r)$ can also be obtained from their published results. Notice that full pyramids have all the same relaxation factor $R(0.5) = 0.45$ whatever their size. Thus they all have the same elastic energy density but obviously have different elastic energies proportional to their volume. The relaxation factor dependence with r so obtained is reported in figure 4c. Again as for the rectangular box there is $R_A(0) = 1$ and $R_B(0) = 0$ since r=0 represents a continuous film. Again $R_A(r)$ drops for increasing r but stops at the limiting value $r_{max}$=0.5 where it has a finite value of 0.35 with a zero derivative. As for the box shaped crystal $R_B(r)$ increases passing a slight maximal value and limiting itself at 0.1 for $r_{max}$=0.5 with a zero derivative. For our analytical use these results can be numerically adjusted by $R(r) = R_A(r) + R_B(r) = 1 - 1.4[r(1-r)]^{0.65}$ giving as it should $\partial R/\partial r|_{r=0.5} = 0$ (see fig 4c).

### III.2/ Equilibrium shape change due to strain

We will separately consider box shaped and truncated pyramids, then we discuss the so obtained quantitative results. In fact we have to explicit the generalised Wulf-Kaishew's theorem (II2 according to (8), (8'), (14))

#### III.2.1/ Box shaped crystal

For a box shaped crystal, $\theta_i = \pi/2$, $n_i$=4, $h = l/2$, there are only two independent surfaces to consider, the interface $S_{AB}$ and the lateral face $S_{//}$ of surface energy $\gamma_{//}$ (see figure 2b) so that the generalised Wulf Kaishew theorem (8), (8') reads:

$$\frac{\Delta\mu - \mathbf{E}_o m^2 R}{2} = \frac{\gamma_{//} + \frac{V}{4}\mathbf{E}_o m^2 \left.\frac{\partial R}{\partial S_{//}}\right|_{S_{AB}}}{l/2} \qquad (18)$$

$$\frac{\Delta\mu - \mathbf{E}_o m^2 R}{2} = \frac{2\gamma_A - \beta + V\mathbf{E}_o m^2 \left.\frac{\partial R}{\partial S_{AB}}\right|_{S_{//}}}{H} \qquad (18')$$

Changing the variables of the partial derivatives to the aspect ratio $r = H/l$, see appendix B, (18), (18') can be resolved parametrically by the parametric representation:

$$\begin{cases} H_{eq} = -\frac{4\gamma_{//}}{3\mathbf{E}_o m^2}\left(1 - \frac{r_o}{r}\right)\left(\frac{dR}{dr}\right)^{-1} \\ l_{eq} = H_{eq}/r \end{cases} \qquad (19)$$

where we put
$$r_o = \frac{2\gamma_A - \beta}{2\gamma_{//}} \qquad (20)$$

the so-called wetting factor. $r_o$ also is the aspect ratio of a strain free deposited crystal (m=0) as can be seen from the ratio of (9') and (9). System (19) only has solutions for $r > r_o$. This condition already means that epitaxially strained box shape crystals have greater aspect ratios $r > r_o$, than strain free crystals for which $r = r_o$. Calculating then $dR/dr$ from fig 4d we can plot $H_{eq} = H_{eq}(l_{eq})$ for a given wetting factor $r_o$ and the $\gamma_{//}/\mathbf{E}_o$ value. Notice that this type of ratio $\gamma/\mathbf{E}_o$ may be called elasto-capillary length. With $\mathbf{E}_0 \approx E$ where E is Young's modulus in an isotropic surface of surface energy γ, this length scales $10\gamma/\mathbf{E}_o \approx 10^{-8}$ cm that means with the size of an atom as mentioned by F.C.Frank[6] [40]. We will use it in the following. The result of the resolution of (19) is shown in figure 5 for $r_o=0.1$, K=1 for m=0 (fig. 5a) and m=4% (fig 5b). Each curve in figures 5 is the trajectory of the edge of the ES

---

[6] We verified that this relation is true around ±30 % for a very great variety of clean surfaces from W, Pt to Si, Ge via true metals and alkali halides up to noble gas crystals. Clearly this ratio is sensitive to foreign adsorption.

with size.

For m=0 the usual Wulf-Kaishew theorem still holds and the equilibrium aspect ratio r=$r_o$ is size independent or the origin of the coordinates, here O, is the self similarity centre we evidenced in figures 3b,3c,3d. For m≠0 the ES ratio increases with size as shown by the two half-crystal shapes plotted on figure 5b for small and large volumes. A more complete discussion will be given in section III.2.3.

**III.2.2/ Truncated pyramids**

For truncated pyramids, using geometrical data of figure 2a, the generalised Wulf Kaishew theorem (8), (8') reads

$$\frac{\Delta\mu - \mathbf{E}_o m^2 R}{2} = \frac{\gamma_1 - \gamma_A \cos\theta + \frac{V}{4}\mathbf{E}_o m^2 \left.\frac{\partial R}{\partial S_1}\right|_{S_{AB}}}{h_1 - h\cos\theta} \quad (21)$$

$$\frac{\Delta\mu - \mathbf{E}_o m^2 R}{2} = \frac{2\gamma_A - \beta + V\mathbf{E}_o m^2 \left.\frac{\partial R}{\partial S_{AB}}\right|_{S_1}}{H} \quad (21')$$

where $\gamma_1 \neq \gamma_A$ is the surface energy of the oblique facet of area $S_1$. Changing the variable to $r = H/l$, see appendix B, the ES can thus be described by the following parametric representation:

$$\begin{cases} H_{eq} = \frac{\left[tg(\alpha/2) - \frac{r'_o}{r}(1 - 2r\cot g\theta)\right](1 - 2r\cot g\theta)}{\left(\mathbf{E}_o m^2 / 2\gamma_A\right) k(r)\left[3/2 - 3r\cot g\theta + 2(r\cot g\theta)^2\right]} \left|\frac{dR}{dr}\right|^{-1} \\ l = H_{eq}/r \end{cases} \quad (22)$$

Where 
$$r'_o = \frac{2\gamma_A - \beta}{2\gamma_A} \quad (23)$$

is the wetting factor and

$$tg(\alpha/2) = \frac{\gamma_1 - \gamma_A \cos\theta}{\gamma_A \sin\theta} \quad\quad (24)$$

In (24) $\alpha$ is the angular extension of the top facet A measured from the Wulf point and for the case m=0 that means obtained from the non strained ES (see figure 4a) or directly from the gamma plot. It is related to the anisotropy ratio $\gamma_1/\gamma_A = \cos\theta + \sin\theta\, tg\,\alpha/2$.

System (22) only has positive solutions H and l for $\frac{r'_o}{tg\,\alpha/2 + 2r'_o \cot g\theta} \leq r < \frac{tg\,\theta}{2}$.

This condition only means that r has to be varied in between a minimum value and a maximum value which is that of a full pyramid. Let us note that for coming back to formula (19) (20) of a box shaped crystal there is to put in (22) $\theta=\pi/2$, from (24) $tg\,\alpha/2 = \gamma_1/\gamma_A \equiv \gamma_{//}/\gamma_A$, since $\gamma_1$ becomes $\gamma_{//}$ substituting (23) in (22). In figures 5c and 5d we plot the equilibrium size $H_{eq} = H_{eq}(l_{eq}/2)$ for K=1, $r_o$=0.1, tg($\alpha$/2)=0.4 respectively for m=0 and m=4%. The ES thus is simply plotted by drawing the oblique facet ($\theta$=45°) for a given lateral size l until the equilibrium height $H(l)$ given by the continuous curve calculated from (22) was reached. Then the top facet is drawn. In figures 5c and 5d we also report the locus of the edge of the horizontal and oblique facet obtained as $\frac{H(r)}{2r} - H(r)$. In figure 5c we see that in absence of misfit (m=0) usual Wulf Kaishew theorem is recovered since all the shapes are self-similar. For m≠0 (see fig. 5d), as previously predicted from the general discussion II2 point (iii), the ES continuously changes with volume. For small volume the ES is a flat truncated pyramid whereas for great volume it gradually becomes a full pyramid. As soon as the full pyramid shape is reached, owing to the fact that the relaxation factor no more varies (see figure 4c) further growing crystals remain full pyramid that means self similar as long as no drop of stress occurs by some plastic relaxation.

**III.2.3/ Discussion**

In this paper we only discuss misfit, wetting and surface energy anisotropy effects. Relative rigidity effects of deposit and substrate and elastic interaction in between crystals have been discussed in [24].

*III.2.3.1 Misfit effect:*

In figures 6a, 6c we plot the equilibrium shapes (ES) calculated from (16) and (22) as trajectories of the edges for box shaped (6a) and pyramids (6c) for different misfit values m. On each trajectory m are drawn shapes for identical volumes. As previously written, for a given volume the equilibrium shapes depend upon the misfit value. Let us note that since we do not consider surface stress effects only $E_o m^2$ appears in the generalised theorem (8), (8') so that the ES *only depends upon the absolute value of the misfit m*. Globally we see on figures 6a, 6c that for zero misfit the ES are self similar for all volumes whereas for non vanishing misfit the greater the misfit m the smaller the interfacial contact area and thus the greater the ES thickening. If for a box shaped crystal (fig. 6a) the misfit effect only is a thickening (depicted by the deviation of the trajectory from the straight trajectory describing the stress free ES), for other shapes the epitaxial misfit also plays a role on oblique facets extension. For instance we plot in figure 6c the epitaxially strained ES of a crystal for which the stress free ES is a severely truncated pyramid. For increasing misfit values the top facet area decreases. We illustrate it in figure 6c for pyramids of equal volume when m=0, 4% and 8%. For zero misfit the trajectory is a straight line (self-similar growth) whose slope is $2r'_o/(tg(\alpha/2)+2r'_o)$. For m=4% the crystal is higher and its top facet area roughly is half that for m=0. For m=8%, since the trajectory intercepts the vertical axis at H≈25, all crystals higher that 25 atomic planes have no more top facet on their ES and thus grow in a self similar way as full pyramids. Let us note that for vanishing lateral size all the trajectories have for asymptote the trajectory calculated for m=0. Obviously the greater the misfit strain, the greater the deviation from this asymptote.

*III2.3.2 Strained ES versus wetting ratio $r_o$:*

In figure 6 we also plot the ES of box shaped crystals (fig 6b) and pyramids (fig 6d) fixing misfit at m=4% but changing the wetting ratio. For box shaped crystals for $r_o$=0.1, $r_o$=0.2 and $r_o$=0.4, according to fig 6b, the deviation from linearity (similarity) is all the more important the wetting factor is small. The same is true for pyramidal crystals for which we plot in figure 6d for $r_o$=0.1, $r_o$=0.2 and $r_o$=0.6 the three trajectories. For the two cases (box shaped, pyramid) the greater the wetting (weak $r_o$) the greater the deviation from the stress free ES. This means that *if the wetting flattens the ES, the epitaxial strain acts against wetting and thus "blows up" the ES and favours oblique facet extension to the detriment of top facet.* It is also important to note that all the pyramidal crystals become full pyramids for the same height *whatever the wetting factor $r_o$*. In figure 6d where we took m=4 % and tg$\alpha$/2=0.4 this height is reached for $H_{FP}$=l/2≈100 in atomic units (a.u) where all trajectories converge. More generally this common height beyond which all growing crystals are full pyramids can be easily calculated by injecting $r_{max}$=1/2tg$\theta$ in equations (22) and using appendix B, knowing that $dR/dr|_{r\max} = 0$:

$$H_{F.P} \approx \frac{12\gamma_A}{\mathsf{E}_o am^2} tg(\alpha/2) \qquad (25)$$

result which is independent of the pyramid angle θ.

Let us note that all these shape changes happen for nanometric sizes (see the above numerical application where $\gamma_A/\mathsf{E}_o a \approx 1/30$, with a the size of an atom).

*III2.3.1 Versus Stranski-Krastanov growth mode*

Until now we only have been concerned with Volmer Weber growth where 3D islands A stay on a bare substrate B since we studied $r_o$>0 that means according to (20),(23) the case $2\gamma_A - \beta > 0$. It is also important to describe the Stranski-Krastanov case where the island A stay on z underlying strained layers of A. In this case owing to the necessary spontaneous formation of the 2D wetting layers the *wetting ratio $r_o$ of A/B must be negative but the wetting ratio of the 3D island A onto the strained wetting layers A vanishes with the number of layers and thus can be taken $r_o=0$* [41,42] [7]. Furthermore, since to our best knowledge the relaxation factor $R_B$ of a composite substrate (z A layers + semi-infinite crystal B) is not available up to now[8], we will consider that the contribution of the 2D layers to the relaxation factor of the substrate $R_B$ can be neglected. It is all the more true the number of these layers is small ( for instance 1≤z≤3 for $Si_xGe_{1-x}$/Si [43]). Thus within these approximations the SK case can be shorthand studied as the limiting case $r_o=0$ in our previous relations.

In figure 7a we plot the equilibrium curve $H_{eq} = f(l_{eq}/2)$ for truncated pyramids θ=π/4 for decreasing parameter $r_o$ ( $r_o=0.2$, $r_o=0.01$, $r_o=0.002$, and $r_o=0$) with current values of tgα/2 =0.4 and m=4%. It is surprising that for vanishing values of the wetting ratio $r_o$ this family of equilibrium curves exhibit an increasing shoulder whose maximum asymptotically tends towards $H_{eq}=0$ and $L_{eq}\to\infty$ for $r_o\to 0$. Thus for a given lateral size $l_{eq}$ there can exist 1,2 or 3 values of the equilibrium heights (see the vertical lines cutting the curves on figure 7a). This peculiar behaviour can be understood by plotting the normalised free energy change $\Delta f = \Delta F/(2\gamma_A V)$ of (5) as a function of the aspect ratio r for given volumes V (see figure 7b). For a truncated pyramid this energy change reads by using formulae (c) and (f) of appendix B with definitions (23) and (24):

$$\Delta f = \frac{E_o m^2}{2\gamma_A}R(r) + [r_o + 2tg\,\alpha/2\,r(1-r\cot g\theta)]V^{-1/3}(rk(r))^{-2/3} - \frac{\Delta\mu}{2\gamma_A V} \qquad (26)$$

---

[7] In fact if the wetting condition is duly expressed in terms of short range interaction amended by long range forces the wetting ratio becomes z dependent $r_o(z)=r_o f(z)$ where f(z) is a quickly vanishing function of z and we take $r_o f(z)=0$.
[8] Green functions for such composite substrate too.

In figure 7b we plot in arbitrary units $\Delta f(r)$ for given V (in atomic units) and $r_o$. In all the cases the normalised chemical potential $\Delta\mu/(2\gamma_A V)$ only vertically shifts the different curves, so our calculations have been made for Δμ=0.

* For $r_o$=0.2 (curves 1 and 2 on fig 7b with respectively V=1.8 $10^4$ and 2 $10^4$) $\Delta f(r)$ exhibits one minimum in the permit range of 1/4<r< 1/2

* For r'$_o$=0.002 the minimum value of r becomes close to zero so that $\Delta f(r)$ behaves very differently. For small enough volume $\Delta f(r)$ only exhibits one minimum for very small values of r that means for flat crystals (see fig7b curve 3 where V= 3.5 $10^4$) . For greater volumes (fig 7b curves 4 and 5 with respectively V= 4.1 $10^4$ and 4.4 $10^4$) $\Delta f(r)$ exhibit three extrema (one maximum, two minima) corresponding to the three possible heights for a given lateral size (in figure 7a see vertical lines with 3 dots). More precisely for a given volume V there are two ES (minima of $\Delta f(r)$), a flat one and a thicker one. For increasing volumes, both aspect ratios increase but furthermore the thickest shape becomes more stable (minimum minimorum of $\Delta f(r)$) so that there can be a shape transition from flat to thick crystal

* For r'$_o$=0, $\Delta f(r)$ looks like the curves reported by Duport et al [28]. For small volumes (fig 7b curve 6 with V=4 $10^4$) $\Delta f(r)$ is minimum for r=0 so that the ES thus must correspond to a 2D layer. For great enough volume (fig 7b curves 7 and 8 with respectively V=7 $10^4$ and V= 8 $10^4$) there is a second minimum corresponding to a thicker crystal. These two minima correspond to the two possible heights seen in fig 7a (stars on the vertical line). A transition from 2D (r→0) to 3D shape can occur at constant chemical potential Δμ as soon as $\Delta f(r) - \Delta f(r=0) < 0$ that means according to (26) for

$$\left(\frac{E_o m^2/2\gamma_A}{2 tg\,\alpha/2}\right) V^{1/3} \geq \left[\frac{r(1-r\cot g\,\theta)}{(1-R(r))(rk(r))^{2/3}}\right] \qquad (27)$$

It clearly appears that the driving force for SK transition is the elastic relaxation since in absence of any relaxation (R=1) only a flat crystal of infinite volume could transform, at constant volume, in 3D crystal !

Since for θ=45° the right side term of (27) has a minimum amounting to 1.43 for r=0.3, the Stranski Krastanov transition only may occur when the volume of the flat crystal fulfils the following relation:

$$V_{SK} \geq \left[ \frac{2.7 tg(\alpha/2)}{\mathsf{E}_o m^2 / 2\gamma_A} \right]^3 \quad (28)$$

which is analogous to the condition of SK transition given by Duport et al [28] but with factor 2.7 in place of π. However calculated activation energies necessary for such a transformation are numerically extremely high. Nevertheless the volume and thus the activation energy necessary for Stranski Krastanov transition could be lowered by decreasing tgα/2 value (foreign adsorption onto the pyramidal faces, roughening transition of the (001) facet…) or by increasing the misfit value. In another paper [42] we have shown how owing to strain relaxation, large enough 2D islands of single heights double their height and start the Stranski-Krastanov transition.

*III2.3.4 Strained ES versus surface energy anisotropy tgα/2:*

In figure 7c we plot the ES trajectories of pyramidal crystals having two different top facets extension tg(α/2) on the stress free ES, $r_o$ and m being the same. For the two ES but of same volume (drawn for tgα/2=0.6, 0.2 or 0) it is seen that the disappearance of the top facet is all the more easy the facet has a less extension on the strain free ES. For a vanishing tg(α/2) ES trajectory is the vertical asymptote (l =0) (see fig 7c). As previously seen from (25) if the stress free ES still is a full pyramid, the epitaxially strained ES also is a full pyramid. In other words, *misfit strain modifies the various facet extension when they exist on the stress free ES but it cannot create new facets that do not exist on the stress free ES*. We will come back to that point in part IV.

**III.3/ Dislocation entrance and shape change:**

We have seen that in all cases epitaxial strain acts against wetting and thus leads to crystal thickening with a decrease of the interfacial and top face areas to the profit of all other oblique facets. Nevertheless during crystal growth the 3D islands, even partially relaxed accumulate strain energy which may become prohibitive. In section II3 we have given the thermodynamical criterion (17) for dislocation entrance and abrupt modifications (in particular flattening) of the ES have been predicted for each dislocation entrance. In the following we illustrate the effect of dislocation entrance on the ES in the Volmer Weber case. The Stranski-Krastanov (SK) case is somewhat similar since dislocation entrance has to produce crystal flattening too. Nevertheless in that case since there exists a minimal misfit-dependent-volume $V_{SK}(m)$ (see (28)) and since each new dislocation entrance (from N to N+1) reduces the misfit from $m_N$ to $m_{N+1}$ (see(16)), there can be $V_{SK}(m_N)< V < V_{SK}(m_{N+1})$ where V is the volume of the supported crystal. Thus if a crystal of volume V having N dislocations can exist in equilibrium conditions ($V_{SK}(m_N)<V$) a crystal of same volume having N+1 dislocations can no more exist in equilibrium conditions ($V < V_{SK}(m_{N+1})$). In this case energy considerations show that the $N+1^{th}$ dislocation entrance produces a transition from the 3D crystal to a flat film. Furthermore for the SK case there is to compare thermodynamical criteria for dislocation entrance in the 3D crystal or in the underlying layers. For all these reasons we will closely analyse the SK case in another paper and here only treat the Volmer Weber case of box shaped and truncated pyramidal crystals.

In both cases the thermodynamical criterion for dislocation entrance (17) reads

$$H_N = \frac{K}{2\pi(1+K)} \frac{1+\ln H_N}{\left(m - \frac{N+1/2}{l_N}\right)k(r)R(r)} \quad (29)$$

where for the box shaped crystal $k(r)=1$ whereas for the truncated pyramids $k(r)$ is a geometrical factor given in appendix B. In relation (29) we put $\chi = H_N$ valid as soon as the distance in between dislocations is greater that the thickness H and where $H_N$ and $l_N$ are the critical sizes (in atomic units) when the number of dislocations passes from N to N+1. In figure 8a (resp.8b) we plot (at the left from origin) the ES $H_{eq}(l_{eq}/2)$ for a box shaped crystal with misfit m=4%, $r_o$=0.1 and K=1 (resp. for truncated pyramids with furthermore tg$\alpha$/2=0.4) We also plot the thermodynamical criterion for a first dislocation (N=0) entrance $H_0(l_0/2)$ parametrically calculated from (26) (dots). This curve has an horizontal asymptote which is nothing else than the usual Matthews criterion [8] for dislocation entrance for 2D films ($l \to \infty$ and R(0)=1 in (29)). For a misfit m=4% this asymptote amounts H=6 monolayers. For finite 3D crystals, $l$ finite, greater thicknesses become possible especially due to the relaxation factor R appearing in (29). The locus of the edge of the ES of a slowly growing crystal may follow the equilibrium curve $H_{eq}(l_{eq}/2)$ valid for m=4% and drawn at the left of figure 8a (resp.8b) for box shaped crystal (resp. truncated pyramids) as a heavy line with an arrow.

*The box shaped crystal* accumulates strain energy until it reaches the size $H_0$=61, $l_0$=155 where the two curves $H_{eq}(l_{eq}/2)$ and $H_0(l_0/2)$ intercept (see figure 8a) so that a first dislocation may thermodynamically enter. If the dislocation effectively enters for this size, according to (16) the misfit passes from m=4% to $m - 1/l_o \approx 3.4\%$. The ES of the growing crystal thus must follow a new ES re-calculated from (19) but with the new misfit 3.4%. How the 3D crystal goes from its former shape ratio to the new one is a matter of kinetics of material transport. In figure 8a we assume that the island changes its shape at constant number of atoms. The crystal thus passes from ($H_o \approx 61$, $l_0 \approx 155$) to a new size ($H \approx 54$, $l \approx 178$) so that the crystal abruptly flattens. The second pair of dislocations enters when the new ES calculated for m=3.4% and the curve giving the thermodynamical criterion for a second dislocation entrance calculated from (29) with N=1 cross each other. This occurs for the size ($H_1 \approx 61$, $l_1 \approx 193$) the misfit then passes from 3.4 % to $m - 2/l_1 \approx 2.9\%$ and thus the growing crystal now follows a new equilibrium curve calculated for 2.9% and so on and so fourth.

*For truncated pyramids* the scenario is somewhat similar. Before the crystal size reaches the value $H_{FP} = l/2 \approx 100$ monolayers (ML) (see (25)) where the ES would become a full pyramid (and then the crystal should grow homothetically following a straight line r=1/2 in figure 8b) the thermodynamical criterion curve $H_0(l_0/2)$ calculated from (30) with N=0 and the ES $H_{eq}(l_{eq}/2)$ intercept so that a first dislocation pair enters at $l_o \approx 152$ (see fig. 8b). Strain thus is released from m=4 % to $m - 1/l_o \approx 3.3$ %. The second pair of dislocations enters for a critical size $l_1 \approx 174$. The strain thus is lowered from 3.3 % to $m - 2/l_1 \approx 2.8$ % and the ES must follow the new equilibrium curve $H_{eq}(l_{eq}/2)$ calculated from (26) with m=2.8 %. The ES again is modified and so on and so fourth ( for N=2, $l_2 \approx 196$ and the residual stress becomes 2.5%, then for N=3, $l_3 \approx 112$ it becomes 2.2 %).

On figure 8a we plot as a fat curve with arrows the equilibrium curve followed by the growing box shaped crystal when 1, 2 then 3 dislocation pairs enter. The remarkable effect is that *for a box shaped crystal the ES ratio drops during each dislocation entrance. Furthermore for pyramidal crystals the top face extension increases at each dislocation entrance*. In figure 8b we sketched the ES of such truncated pyramids just before the first and just after the third dislocation entrance. The greater the volume, the greater the dislocation number and thus the flatter the crystal and the greater the top face extension. Let us note that for high enough misfit the first dislocations enter in full pyramid, then for a great enough number of dislocations the top face reappears on the ES.

# IV/ Connection to other works and discussion

## IV.1/ Relaxation factor and ES

We have seen that the difference in between the strain free ES and the epitaxial strained ES comes from the elastic energy stored by the epitaxial system $E_o m^2 VR$ (see (3)). The ES thus depends upon the partial space derivatives of the relaxation factor R (see (11)). Thus the more accurate the calculation of the relaxation factor R as a function of shape, the more accurate the theoretical prediction about ES change. In literature R factors have been calculated in some particular cases we want now discuss.

### IV.1.1/ Analytical model calculations

Analytical calculations of the relaxation factor R are complex and need an accurate description of the elastic interaction in between the deposited crystal A and its substrate B. Some authors [28,29,44] modelise this interaction by a 2D a priori distribution of point forces representing an elementary action of A on the interfacial area in between A and B. In all cases then the Green elastic tensor is used to derive stress then strain and consequently the elastic energy stored in the substrate.

* Tersoff [29] describes the epitaxial contact by a distribution of point forces

$$f_j = \frac{\partial}{\partial x_j}[\sigma h(x)] \approx \sigma \frac{\partial h(x)}{\partial x_j} \qquad (31)$$

where σ is the in plane stress supposed constant in all the island volume and h(x) the height of the island at position x. By this way, the first Tersoff' approximation is to neglect the stress change in the deposit $h\frac{\partial \sigma}{\partial x}$ precisely due to the elastic island relaxation.

For a box shaped crystal such distribution of forces $f_j$ of (31) becomes *a distribution of elastic monopoles* $\pm \sigma h \delta(x \pm l/2)$ located at the edges $x = \pm l/2$ of the box. Let us note that Hu [45] stated long before that the concept of such concentrated edge forces is not sound for describing epitaxial contacts: the stress change $h\frac{\partial \sigma}{\partial x}$ during the island relaxation cannot be neglected since if the substrate becomes deformed by such epitaxial forces, thus in turn it deforms the deposited crystal and thus leads to a new force distribution. The distribution of epitaxial forces thus has to be determined by a self consistent analysis as we above mentioned and used in sections III.1 and III.2 following [38].

For a truncated pyramidal crystal of angle θ the interfacial distribution of point forces (31) of Tersoff becomes $f_x = \sigma tg\theta$ for $-l/2 < x < -l/2 + h\cotg\theta$, $f_x = -\sigma tg\theta$ for $l/2 - h\cotg\theta < x < l/2 + h\cotg\theta$, and $f_x = 0$ anywhere else. This means that only the part of the interface where $f_x$ differs from zero plays a role in the substrate striction effect. This is not physical for epitaxial contacts ! Once more, distributed forces have to be calculated in a self-consistent way. Nevertheless, using such modelised epitaxial forces, and then using the Green tensor for isotropic crystals, Tersoff [29] finds an analytical expression for the elastic energy stored in the substrate B: $\mathsf{E}_o m^2 V^A R_{S=B}(r)$ with $R_S(r) = r\ln(e^{3/2} tg\theta / r) > 0$ within the (second) approximation of severely truncated pyramids, $r << r_{\max} = tg\theta/2$. If we strictly follow Tersoff argument, the total elastic energy the system (A+B) stored is $W_{el} = \mathsf{E}_o m^2 V^A (R_A(r) + R_{S=B}(r))$ with $R_A(r) = 1$ since in the Tersoff first approximation the deposit relaxation is neglected..

In fact Tersoff as well as the other authors studied the so-called Stranski-Krastanov (SK) case where the 3D deposit A is coherently bond to a composite substrate with z *prestrained layers* of A ($\varepsilon_{xx} = \varepsilon_{yy} = m$) supported by a semi infinite matched substrate B. In such case the force distribution (31) of the 3D crystal locally *stresses back* the film so that the elastic energy in the film drops from the elastic energy $W_{lay.}$ stored by the z pseudomorphous layers before any relaxation to $W_{lay} - \mathsf{E}_o m^2 V^A R_{S=A}(r)$ after relaxation where $R_{S=A}(r)$ again is the above-mentioned logarithmic form[9]. The elastic energy of the system is now $W_{el,SK} = W_{lay} + E_o m^2 V^A (R_A(r) - R_{S=A}(r))$ where, in the framework of Tersoff first approximation, there is again $R_A(r) = 1$. This therefore brings a quantitative misbalance but physically makes still some qualitative sense, for the SK case as we will see now. Consider the transformation of a SK film having z' layers (decomposing into z layers of elastic energy $W_{lay.}$ and (z'-z) layers) into a 3D relaxed crystal of volume $V^A$ staying on z underlying layers. Before transformation there is $W_1 = W_{lay} + E_o m^2 V^A$ where $E_o m^2 V^A$ is the elastic energy of the (z'-z) layers that will be transformed in a 3D crystal A of volume $V_A$. After transformation and then elastic relaxation the elastic energy is $W_2 = W_{el,SK}$ so that the energy change during the transformation reads $\Delta W_{1,2} = W_2 - W_1 = E_o m^2 V^A (R_A(r) - R_{S=A}(r) - 1)$ that is the thermodynamic driving force for SK transition. Tersoff first approximation $R_A(r) = 1$ gives $\Delta W_{1,2} = -E_o m^2 V^A R_{S=A}(r) < 0$ with $R_{S=A}(r) = r \ln(e^{3/2} tg\theta / r)$ that means that owing to its negative sign, this force is able to drive the SK transition what these authors wanted to show.

These authors had no special interest in ES changes, however their formula (6) based on the above relaxation factor gives a lateral shape dependence s with the height H that can be written by the parametric representation $H(r) = \frac{\Gamma}{C} \left| \frac{dR_{S=A}}{dr} \right|^{-1}$, $s(r) = H/r$ which is nothing else that our general solution (22) where one can identify for the SK case (r'$_o$=0) with the

---

[9]This result stays correct as long as the thickness z of the SK layers is great ($z > l$ may be enough) so that at the A/B interface the induced displacement field can be neglected. If this is not the case there is created in the otherwise stress free substrate B stress and strain giving a supplementary positive term $E_o m^2 V^A R_{S=B}(z,r)$.

Unfortunately at our best knowledge Green' elastic tensor components have not been calculated for semi infinite planar bicrystals A/B. Even finite elements calculations of that type are not available.

material constants $\Gamma = 4\gamma_A tg\,\alpha/2$, $C = 3E_o m^2$. However due to the condition r<<$r_{max}$=1/2tgθ the resolution is quite limited to very truncated pyramids so that shape changes can hardly be put in evidence.

* Duport et al [28,44] used similar model calculations with point forces but they could extend their result up to full pyramids 0<r< ½ tgθ. However the same criticisms (lack of self-consistency) can be addressed. Here the elastic interaction between deposit and substrate is modelised by a 3D array of discrete *elastic dipoles* in the deposit. For pyramids these dipoles are distributed as atoms in basal layers so that the pyramidal faces become stepped faces. The dipoles are then projected onto the interface where they produce displacements in the substrate storing thus elastic energy. *Since their elastic dipoles, derive from point forces, uniform sheets of dipoles are equivalent to monopoles at the borders of the sheet*. When comparing these results and that of Tersoff [29] in their very restricted common field of validity, r<< ½ tgθ and tgθ<1/4 that means 0<r<<1/8, the same analytical form of the relaxation factor $R(r) = -r\ln\eta/r$ is obtained [28]. Owing to the respective use of multiple integrations and summations, some numerical but minor differences appear for $\eta$.

**IV.1.2/ Finite elements calculations**

Most authors who numerically compute elastic energy due to misfit strain use Finite Element programs of continuous elasticity. The misfit strain is simulated by assigning different coefficients of thermal expansion to the deposit and substrate and then change the temperature of the system so that the interface becomes coherent [26,39,46,47]. It is important to note that in these numerical calculations the so calculated relaxation factor R includes the contribution of the deposit $R_A$ as well as the contribution of the substrate $R_B$. Generally $R_B$ is not distinguished so that is only $R = R_A + R_B$ is available.

S.Christiansen et al [46] have probably been the first to use such 3D methods for polyedral crystals, especially on square based truncated pyramids having various slopes θ.

Unfortunately from their paper it is not clear how the misfit reduction they define is connected to the energy relaxation factor $R(r)$. Furthermore some doubts also appear whether the energy calculations have been done for the deposit crystal only or also for the substrate.

E. Pehlke [47] has specifically studied the strongly strained (7%) InAs:GaAs(001) system. For a non-supported crystal the ES contains (at least) the four stables faces (cusps in the gamma plot) with the decreasing importance ½ {111}, ½ {100}, {110} forms. For this strained crystals they use continuous linear elasticity by a finite element method where the total elastic energy $W_A+W_B$ (no SK underlying layers are considered since thin) is directly minimised in respect to the displacement field. These calculations have not been repeated for each shape and size but scaling laws proportional to volume V for elasticity and proportional to $V^{2/3}$ for surface have been used. At the end of their paper the scaling law in V for elasticity was declared inadequate. A detailed study of a truncated pyramid θ=45° has been done as a function of shape ratio r. For a full pyramid a total energy relaxation factor $R=R_A+R_B=0.35$ not too far from R(1/2)=0.45 of Ponchet' study [39] has been found.

Likely the most accurate and comprehensive R factor calculations have been done by D. Wong et al [27]. After Freund et al [26] having studied the cylindrical 2D case they studied spheres of A coherently mismatched on a flat substrate B. Using finite element program in the framework of continuous linear and isotropic elasticity with proper boundaries conditions, they calculated the total (deposit + substrate) energy relaxation factor of spheres R(Θ) as a function of a contact angle Θ. The R(Θ) curves, 0< Θ<180° are similar to that one R of figure 4d, thus steeper the substrate is weak. For K=1 and a half sphere (Θ= 90°, r=1/2) they found R=0.1 illustrating the strong drop in respect of the initial strain energy (≈90%) which is a bit greater than that of the equivalent (r=1/2) box shaped crystal (≈84%), see figure 4d.

The main result of their ES study is (similar to that of Freund [26]) that the contact angle increases with misfit and size. *However there was applied the constrain that the shape* *remains a spherical ca*p. The authors [27] consider that this hypothesis is valid to a first order, shape deviations from the sphere being said of the second order. Nevertheless from our calculations we have seen that, owing to strain, ES no more obeys to self-similarity and this is not a matter of a second order effect, it strongly depends on misfit. There is no reason that spheres escape that rule.

### IV.2/ Experimental results

Before comparing our theoretical results with experimental results, for easiness of discussion, let us first *enumerate the main elastic effects on the ES we find*:

*(i) Non self-similarity is the rule and this effect is thus stronger the misfit m is high.*

*(ii) Shape ratios increase continuously with size and the oblique facets increase to the detriment of the top face as long as the crystal remains coherent. Shape ratios for pyramids of angle $\theta$ become maximal for $r = r_{max} = \frac{1}{2} tg\theta$ then they increase self similarly.*

*(iii) At some critical size $L_c$ and some shape ratio $r_c$, dislocations may introduce and then the reverse process of (ii) occurs but in a jerky manner.*

*(iv) The study is restricted to 0 K shapes that means polyhedra. The faces, which suffer the described changes by epitaxial strain, are only those corresponding to inward cusps of the gamma plot.*

*(v) Since surface energy is very sensitive to adsorption the corresponding modifications of the gamma plot have to be considered.*

*(vi) We have not considered strain-induced changes of the gamma plot, that means surface stress effects. Such effects have surely to be considered.*

Since 1990, many experiments on morphology and strain have been done. It is especially the case of semiconductors. Nevertheless most descriptions are incomplete, in respect to the knowledge of: (1) equilibrium or not (2), coherent or dislocated interface, (3) thermodynamic data are missing for full comparison with theory. However we will describe some pertinent experimental evidences reported for several systems. We will write in between brackets the above enumerated points (i) to (vi) at which experimental results can be connected.

*IV2.1 Ge/Si:*

Most experimental data concern Ge on Si (m=-3.6%) where beyond roughly a few Stranski Krastanov layers, large enough 2D Ge islands may thicken and transform in 3D islands. Some experimental evidences about ES have been reported for Ge/Si(111) as well as for Ge/Si(100).

*Ge/Si(111):* On Si(111) initially appear frustum of tetrahedrons {113} with a triangular top face (111) [48]. Voïgtlander et al [49] report the aspect ratio change with coverage ($\propto$ size) observed by in situ MBE-STEM. After 2D island thickening the smallest 3D islands suddenly exhibit a shape ratio of r=0.10. This shape ratio increases then continuously with size (see (i) and(ii)) up to a maximal value r=0.14 for a nano-size L=23 nm (smaller than that of a full pyramid ($r_{Py}$=0.16)). From that size the ratio decreases asymptotically towards r=0.10. This behaviour is associated by the authors [49] with the introduction of dislocations (see (iii)), so that Lc≈23nm is the critical size (what is confirmed by [48] who find all crystals relaxed for sizes 30<L<300nm). The initial appearance and evolution of the {113}(111) pyramids with a dihedral angle of 29° instead of {001}(111) 54° or {111}(111) 70° pyramids can only be understood if one considers that the strain free crystal still contains the dominant flat faces (111) and (113) with cusps in its gamma plot according to (iv). If one accepts some similar behaviour for germanium and silicon, this is true since Bermond et al. [50,51] ES measurements precisely find these faces on silicon under UHV conditions. At their experimental temperature (1300K), {001}{110} faces still passed the roughening transition but are present on the ES as rounded faces.

*Ge/Si(001):* On Si(001), large enough germanium 2D islands, on some two strained monolayers, thicken and transform into 3D crystals. [52]. The square base truncated pyramids {113}(001), with a dihedral angle of ≈25° form first dislocation free [53]. The shape ratio initially close to zero as observed by AFM in-situ [52] continuously goes to that of a full pyramid $r_{Py}$=0.25 (see (i), (ii)) with the nano-size $L_{Py} \approx 100$ nm, then stays constant up to a critical size $L_c \approx 300$ nm where again they become truncated (see (iii)). The shape ratio then decreases hyperbolically with size, the height staying constant ($H_c \approx 60$ nm). Eaglesham et al. [53] changed the ES by additives and post annealing. Clearly the gamma plot has been changed by these additives (see (v)) some existing cusps i having been deepened by selective adsorption on the corresponding faces i.

Remarkable again are the in-situ UHV-TEM studies by LPCVD of Hammar et al [12], confirming but at higher temperature (650°C) the 3D dislocation free growth up to $L_c \approx 100$ nm size, imaging the stress field and by REM showing that the crystals have flat pyramidal {113}(001) shape [53]. Among this size these crystals relax statistically by entrance of misfit dislocations entering from the top and gliding to the interface at some 20nm from the edge. *At each dislocation entrance the crystal edge moves outside by $\Delta L \approx 20$ nm within a few seconds what the authors* [12] *interpreted as a flattening of the shape at quasi-constant volume*. We described this saw tooth behaviour (iii) for $\theta=45°$ pyramids. Again, as for the Si (111) substrate, the ES considerations (point (iv)) are valid explaining why appear on (001) the flat {113} pyramids (25°) and not {011} (45°) or {111}(54°) nor {100}(90°). Let us note that Hansson [54] find from LPE the {111}(001) 54° pyramids what means that (113) faces are no more on the ES when in contact with the liquid phase.

*IV2.2 "huts":*

Up to now we could confirm by experiments our points excepted point (vi). We think that the very special behaviour of the Si(001) face, in respect to germanium deposition at low temperature, called: "huts" formation [55,56,12] belongs to point (vi) of this section. At 350 °C on the strained Ge(001) underlying layers, form very densely packed, square or rectangular and very flat pyramids ($\theta=11°$) with well defined {105} faces. These faces seen by high resolution STEM [55] do not show the usual vicinal structure with steps at every fifth terrace atom. They are "atomically flat" with a herring bone structure 2x1 (105) of dimers [57]. At their optimal size, (L=100 nm with a shallow top face (001) [12]) huts probably are fully strained but do not relax by misfit dislocations. By further deposition, in between them the usual {113}(001) pyramids form and relax at $L_c \approx$ 100 nm by two types of dislocations according to temperature. Huts {015} have also been produced at misfits as small as m=0.8 % from $Si_{0.75}Ge_{0.25}$ layers [58]. The puzzling problem is: *{105} faces do not exist on the strain free ES of silicon* [50,51] and probably not on germanium too. Nevertheless since on non relaxed Ge pyramids {113}(001) growing on Ge strained layers, Tomitori et al [57] found on the basis of these pyramids, small but steep {150}($\theta$=79°) faces, they conclude that "*epitaxial strain stabilises {105}faces beside the {113} one*". In fact we like to go further by the statement: *epitaxial strain can create in the gamma plot new cusps* (here pointing towards the i=<150> directions). The decrease of surface energy at the cusp may be estimated to $\Delta \gamma(i) \approx (s_o - \gamma_o)m$ where $\gamma_o$ and $s_o$ are the surface energy and surface stress in absence of epitaxial strain respectively. Since for clean material there is usually $s_o > \gamma_o$, a cusp creation would happen only for systems where m<0 what is realised for Ge:Si but not for the reverse Si/Ge where m>0. Unfortunately there is an information hole about Volmer Weber growth of silicon on bulk (001) Ge. Let us stress that in another paper Tomitori et al [59] partially bring a strong experimental argument to the above statement. They cut a silicon wafer in (015) orientation and proceed of Ge deposition as a strained film. The observed growth mechanism is that of layer by layer up to 10 to 15 monolayers, each layer showing the 2x1 reconstruction with the same 2x1 reconstructions as the {105} faces of the huts grown on (001) Si. At 300°C the growth mechanism is by birth and spread of 2D islands, at 400° C step flow starts. The steps are those, 0.5Å high, due to usual miscuts <±1° away for the (015) Si. Multiple step heights (2 and 3) occur too. A full proof would be that such a strained (015) film when becoming strain free by some appropriate mean should loose its 2x1 reconstruction or even would break up in facets most probably in skew pyramids {113} (faceting). This problem of "huts" is far being understood. It has to be connected to the more general open question: what drives a surface reconstruction ? More precisely here how external strain may transform a specific vicinal face into a flat singular face ?

*IV2.3 III-V compounds and other materials:*

On the III-V semiconductors similar morphological changes confirming points (i) to (iii), especially "flattening" by dislocation entrance have bee described [60] but the results are much less precise. {105} "huts" or "quantum dots" have been described especially on the high misfit samples of InAs/GaAs(001). Also on other systems where m<0 [13,60] similar facts have been reported. When comparing with the Si-Ge case, things are here experimentally much more complex. Indeed owing to the volatility of the various elements, self-surfactant effects are unavoidable. They interfere with the strain induced shape effects. This is one of the reasons why in the theoretical study on InAs/GaAs(001) [47] the chemical potential of arsenic has been kept constant.

Similar ES studies have been partially reported for Pd/MoS$_2$(0001). In this case Pd crystals grow in epitaxial contact (111) with (0001) of molybdenite [61], those smaller than 15 nm (mean value) being coherently oriented $a<1\bar{1}1>Pd // b<11\bar{2}0>MoS_2$ with a misfit m=(b-a)/a=9 %. Their shapes are triangular based tetrahedrons {111}(001), θ=48° with a top face (001). The shape ratio $r = H/L$ was sampled with size by means of STM by Perrot [62]. It increases from r=0.30 at the size L=5nm up to r=0.42 for the size L$_c$=9.5 nm (see(i), (ii)) where then the height stays constant at H$_c$≈4 nm (full tetrahedron r=0.81). We interpret this quasi-continuous flattening by dislocation entrances followed experimentally up to L=16nm (see(iii)). From TEM study [61] it is clear that the Pd crystals lose at some L$_c$≈10 nm their coherency. They sweep out by ±3° from their original epitaxial orientation, seen by the appearance of double diffraction spots being the result of the introduction of non explicit dislocations [17,18,20].

**V/ Conclusion**

Our generalised Wulf-Kaishew theorem, in its most general form, describes the influence of epitaxial strain on the ES. One of the most striking result is that the ES is no more self-similar with size as it is when no epitaxial strain exists. Have to be accounted the surface energy plot, adhesion energy and bulk elasticity. We left away the effect of elasticity on the surface quantities but discussed (IV22) its presumable incidence in respect to the appearance of "huts" that means faces that do not exist on the strain free ES. Our generalised theorem has been recently amended [63] considering both epitaxial and surface stress effects.

Here ES have been explicitly constructed for a family of shapes ($\gamma$-plots) when the energy relaxation factors R have been available. We could find qualitative agreement with experiments, especially in the field of semi-conductor systems where a great amount of work has been done for technological reasons. However in the future it will be necessary to perform more accurate experiments about ES of epitaxial strained and non strained crystals in order to study quantitatively the aspect ratio change with size. Remarkable is the fact that, though the strain free ES of Si is well known [50,51] the ES of the Volmer Weber system Si/Ge has been completely neglected, whereas at the contrary the ES of the Stranski Krastanov system Ge/Si has been scrutinised in absence of data on the strain free ES of Ge !Such uncomfortable situations are typical for most epitaxial systems where our knowledge on one partner may be acceptable but not for the other partner.

On an other hand, more knowledge has to be accumulated about the calculation of elastic energy relaxation factors R. Various shapes have to be studied, truncated pyramids of different angles $\theta$, more complex forms as pyramids needing the definition of more than one shape ratio e.g. for the cuboctaedra. Last but no least it was not possible to treat in satisfactory way the Stranski-Krastanov mode. It is due to some unsolved problems in respect to the underlying pseudomorphous layers (see III2.3.1). From that analysis (see III3) appears even the doubt about the real existence of an ES window especially when dislocation entrance is considered.

# Appendix A

Let us take any of the cases in figures 3bd and any of the sectors when from S the faces (i) are seen by an solid angle $\alpha_i$ (see fig 9). To show that $(\alpha_i)$ are invariant it is enough to consider one corner or edge e and to prove that the angle $\theta_e$ is invariant. From Wulf point O is measured the distance $h_i$ to the face i and the distance $h_2$ to the next facet (2) noticing that the mutual normals are measured by $\theta_i$ which is a crystallographic invariant. Now just draw points e', e'' and e''' what needs no further explanations from figure 9. It follows

$$tg\,\theta_e = \frac{ee'}{0e''} = \frac{ee''-h_{AB}}{h_2} \quad \text{and} \quad ee'' = (Oe'''-h_2)tg(\pi/2-\theta_i) \quad \text{with} \quad Oe''' = h_i/\sin\theta_i \quad \text{so that}$$

$$tg\,\theta_e = \frac{h_i}{h_2}\frac{1}{\sin\theta_i} - \cotg\theta_i - \frac{h_{AB}}{h_2}$$

By the properties of similarity of the central distances (12), by that one of shifting of Wulf' point (13), by constancy of $\theta_i$, $\theta_e$ thus is invariant.

# Appendix B

\* *For a box shaped crystal* (see fig 4b) simple mathematical transformation gives with $r = H/l$ :

$$\left.\frac{\partial R}{\partial S_{AB}}\right|_{S_{//}} = -\frac{r^2}{lH}\frac{dR}{dr} \qquad (a)$$

$$\left.\frac{\partial R}{\partial S_{//}}\right|_{S_{AB}} = \frac{1}{l^2}\frac{dR}{dr} \qquad (b)$$

\* *For truncated pyramids* (see fig 2 and 4a), simple geometrical considerations and simple mathematical transformations give:

$$V = \frac{H^3}{r^2}k(r) \text{ with } k(r) = 1 - 2r\cotg\theta + (4/3)(r\cotg\theta)^2 \qquad (c)$$

$$S_A = S_{AB}(1-2r)^2 \qquad (d)$$

$$h_1 - h_A \cos\theta = \frac{H}{2r}(1 - 2r\cotg\theta)\sin\theta \qquad (e)$$

$$S_1 = \frac{S_{AB}}{\sin\theta}r(1 - r\cotg\theta) \qquad (f)$$

$$\left.\frac{\partial R}{\partial S_{AB}}\right|_{S_1} = -\frac{r^3}{H^2}\frac{1 - r\cotg\theta}{1 - 2r\cotg\theta}\frac{dR}{dr} \qquad (g)$$

$$\left.\frac{\partial R}{\partial S_1}\right|_{S_{AB}} = \frac{r^2\sin\theta}{H^2(1 - 2r\cotg\theta)}\frac{dR}{dr} \qquad (h)$$

# Figure captions

**Figure 1:** Schematic thermodynamic process of formation of a coherent epitaxial crystal A on a lattice-mismatched substrate B. **1-** formation, **2-** homogeneous deformation for accommodation then adhesion, **3-** inhomogeneous relaxation

**Figure 2:** **(a)** Polyedral crystal A deposited on a substrate B. $h_i$ and $h_{AB}$ are the normal distances from surfaces i (surface energy $\gamma_i$, angle $\theta i$) and from the interface AB, to the Wulf point O. The dotted part of the crystal is the truncated part. **(b)** An extension of the interfacial area increases the elastic energy so that $\partial R/\partial S_{AB}\big|_{S_1,Si} >0$. **(c)** An extension of the oblique facet area decreases the elastic energy so that $\partial R/\partial S_1\big|_{SAB,Si} <0$.

**Figure 3:** ES of increasing volume for m=0. **(a)** For a free crystal ES are self-similar with a common Wulf point. For supported crystals the Wulf point moves (O→O'→O''). For $\beta-\gamma_A>0$ it moves above the interface **(b)**; for $\beta-\gamma_A>0$ it moves beneath the interface **(c)**; for $\beta=\gamma_A$ it stays at the interface (O=O'=O''=S **(d)**. In these cases there is again self-similarity but from S.

**Figure 4:** Gamma plot and connected shapes (height H, lateral length l ) for truncated pyramids **(a)** and box shaped crystal **(b)**. The corresponding relaxation factors, $R_A$, $R_B$ and $R=R_A+R_B$ are given in **(c)** and **(d)**. They have been obtained from [39] and [42] respectively.

**Figure 5:** Half ES (H and l in atomic units) calculated for $r_o=0.1$ Continuous curves are the edges trajectories. *Box shaped crystal*: **(a)** self similar ES for m=0; **(b)** ES thickening for m=4% (dotted line is the trajectory for m=0). *Pyramidal crystal* (with tg($\alpha$/2)=0.4): **(c)** self similar ES for m=0; **(d)** ES change from truncated pyramid to full pyramid for m=4%. (dotted lines are H(l /2) calculated from (22)).

**Figure 6:** Misfit and wetting effect on the ES. For box shaped crystal ($r_o=0.1$) the greater the misfit, the thicker the ES **(a)**. For pyramidal crystal ($r_o=0.1$, tg($\alpha$/2)=0.4) the greater the misfit, the thicker the ES and the smaller the top face **(c)**. In both cases (m=4%), box **(b)** and pyramid **(d)** the greater the wetting the greater the deviation from the stress free ES.

**Figure 7:** **(a)** Edge trajectories of pyramids for $r_o \to 0$ (m=4%, tg($\alpha$/2)=0.4). **(b)** Free energy change $\Delta f(r)$ of (26) for various pyramidal volumes V (atomic units). *For $r_o=0.2$*: *1*,V=1.8 $10^4$; *2*,V=1.8 $10^4$; *For $r_o=0.002$*: *3*,V=3.5 $10^4$: *4*,V=1.8 $10^4$; *5*,V=4.4 $10^4$; *For $r_o=0$*: *6*,V=4 $10^4$; *7*,V=7 $10^4$; *8*,V=8 $10^4$. **(c)** Edge trajectories and ES (of same V) of pyramidal crystals for various values of tg($\alpha$/2)

**Figure 8:** Effect of dislocation entrance. **(a)** Edge trajectories (continuous curves) of box shaped crystals ($r_o=0.1$). From left to right there is m=4%, m=3.4%, m=2.9%, m=2.6% corresponding to crystals having 0,1,2,3 then 4 dislocations. Dotted curves are thermodynamic criterion for dislocation entrance (N=0.1,2,3). Half ES are reported just before the first and just after the third dislocation entrance. **(b)** the same thing for pyramid ($r_o=0.1$, tg($\alpha$/2)=0.4) where from left to right m=4%, m=3.3%, m=2.8%; m=2.5%, m=2.2%.

**Figure 9**: geometric construction (dots) of the ES (fat) from Wulf point O using the distance

$h_i$ normal to facets i.

Figure 1

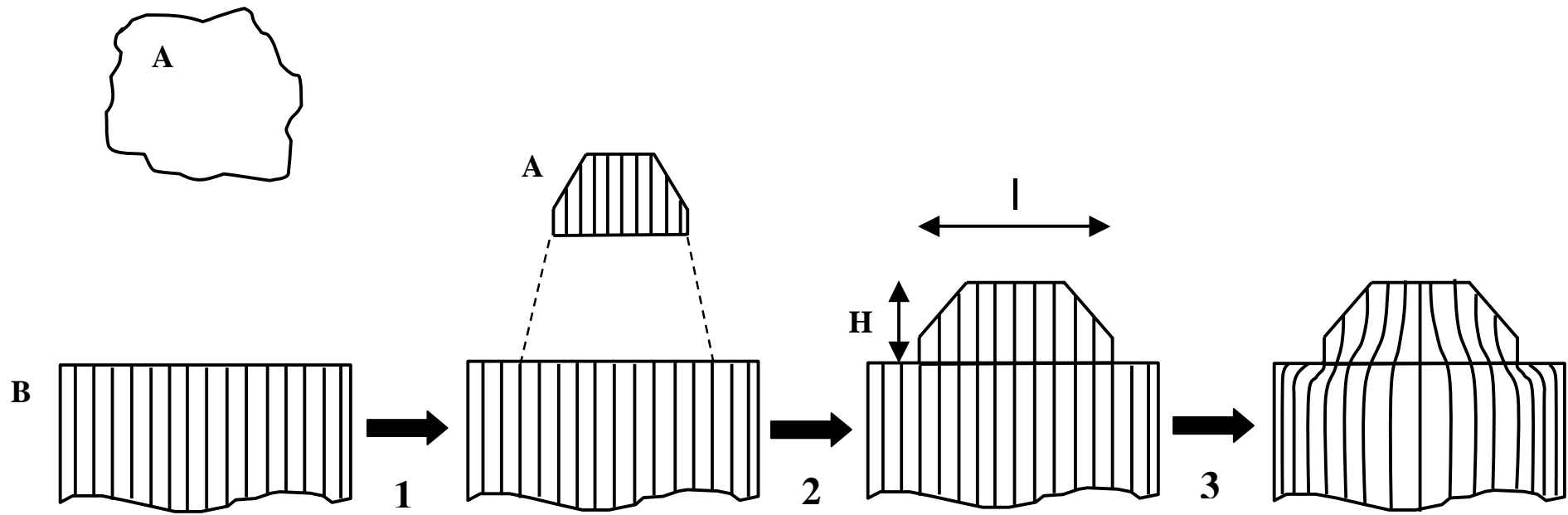

Figure 2

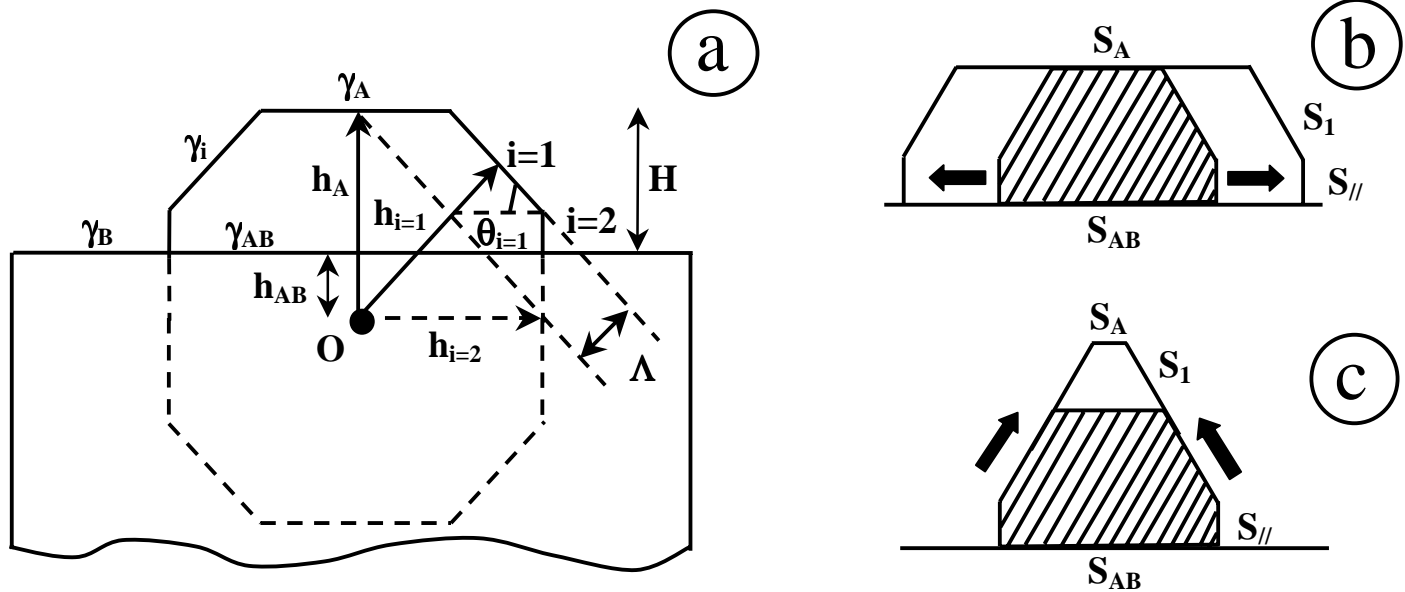

Figure 3 :

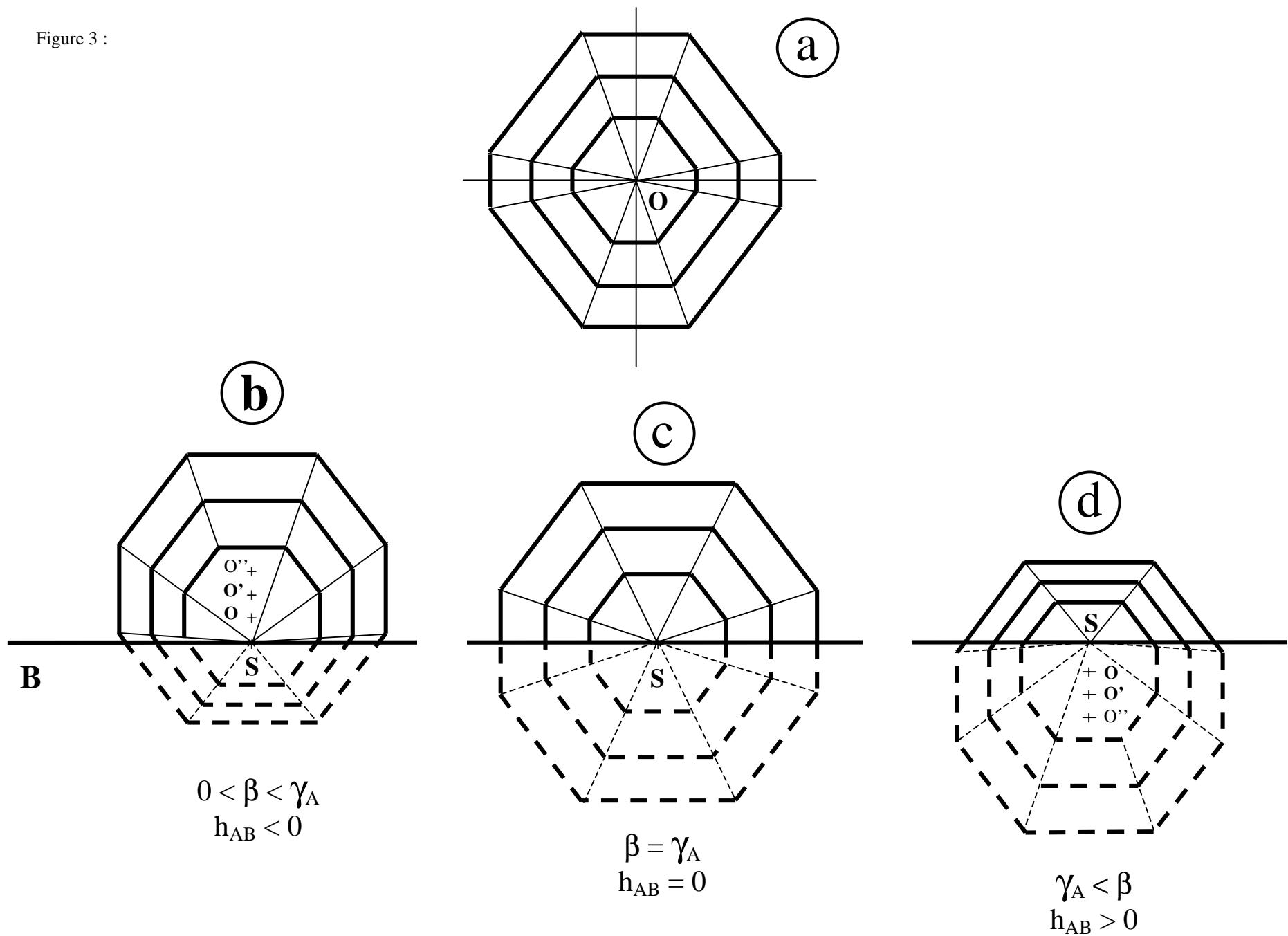

Figure 4 :

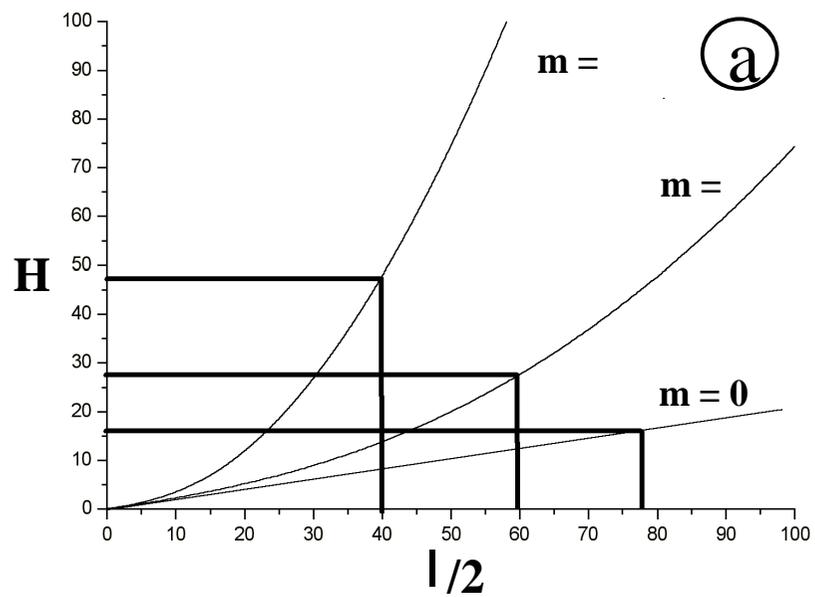
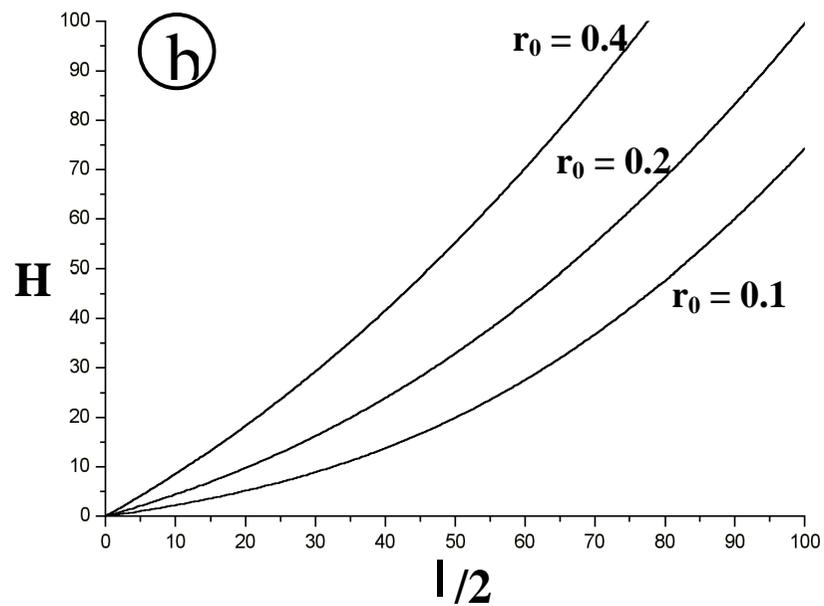
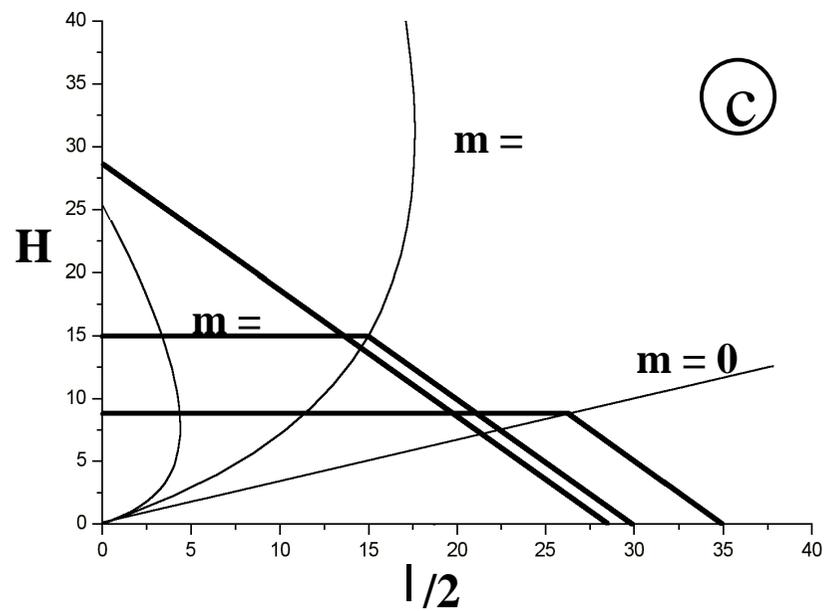
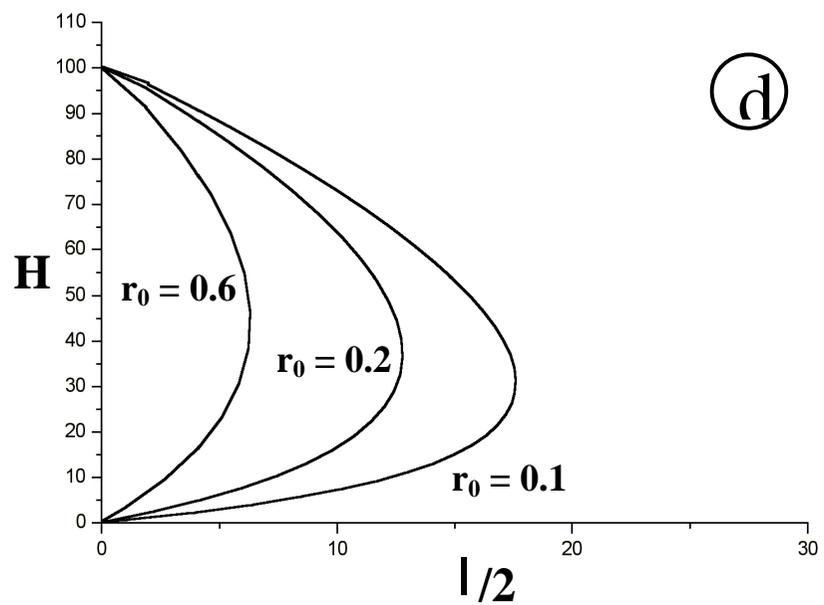

Figure 5 :

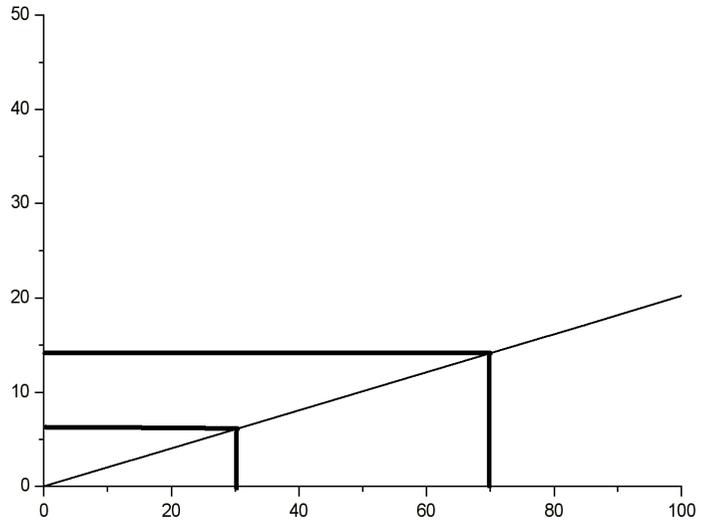
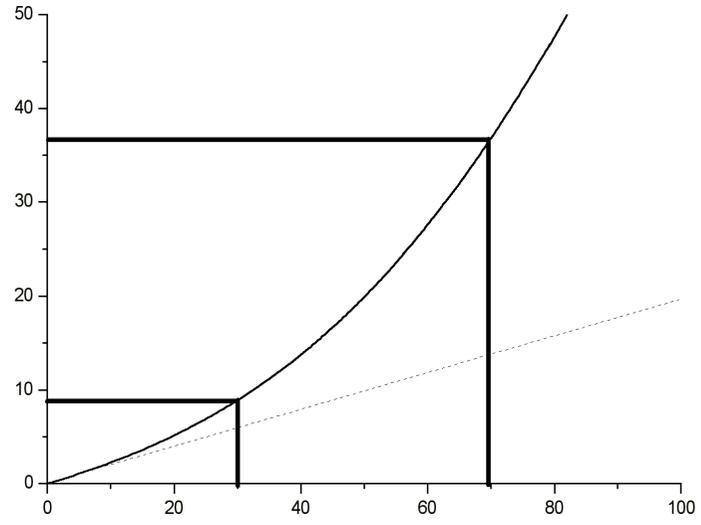
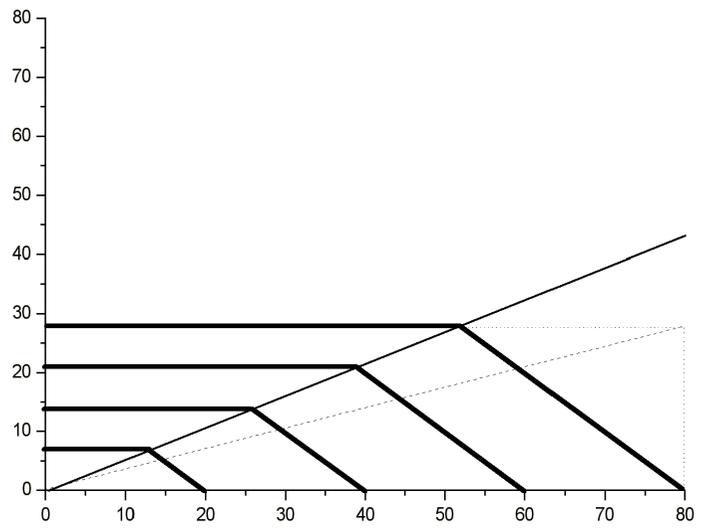
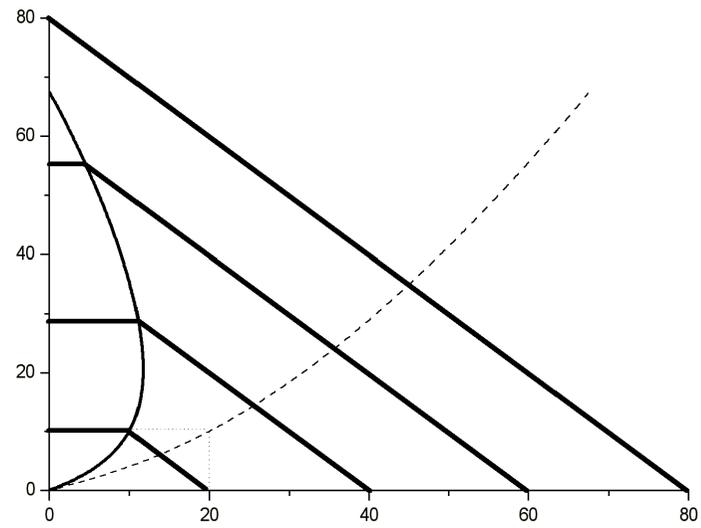

Figure 6 :

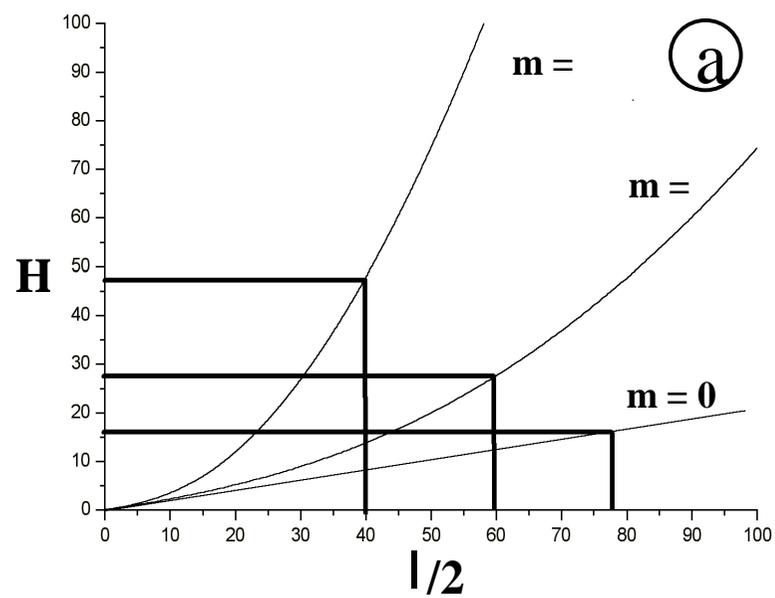
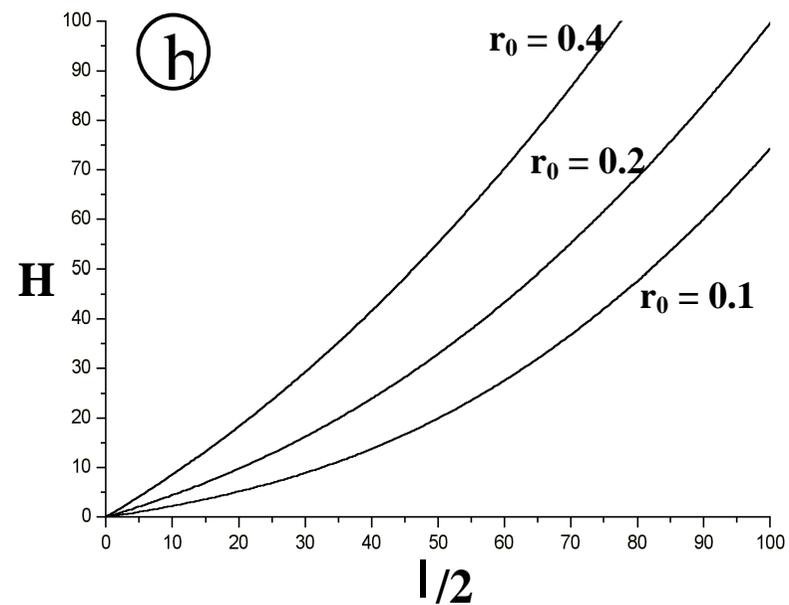
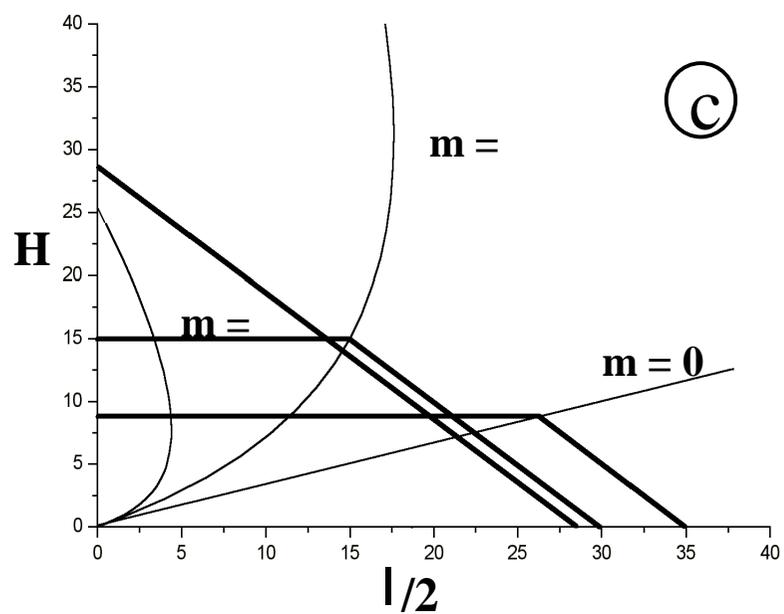
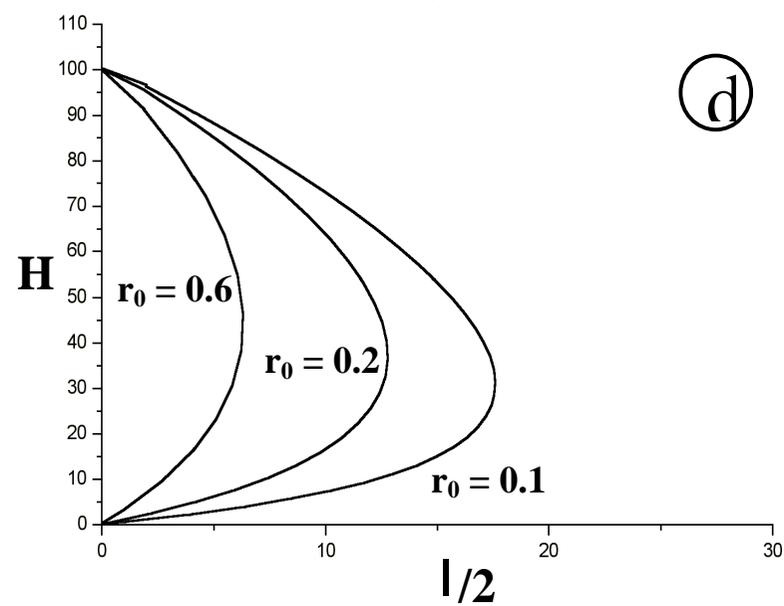

Figure 7 :

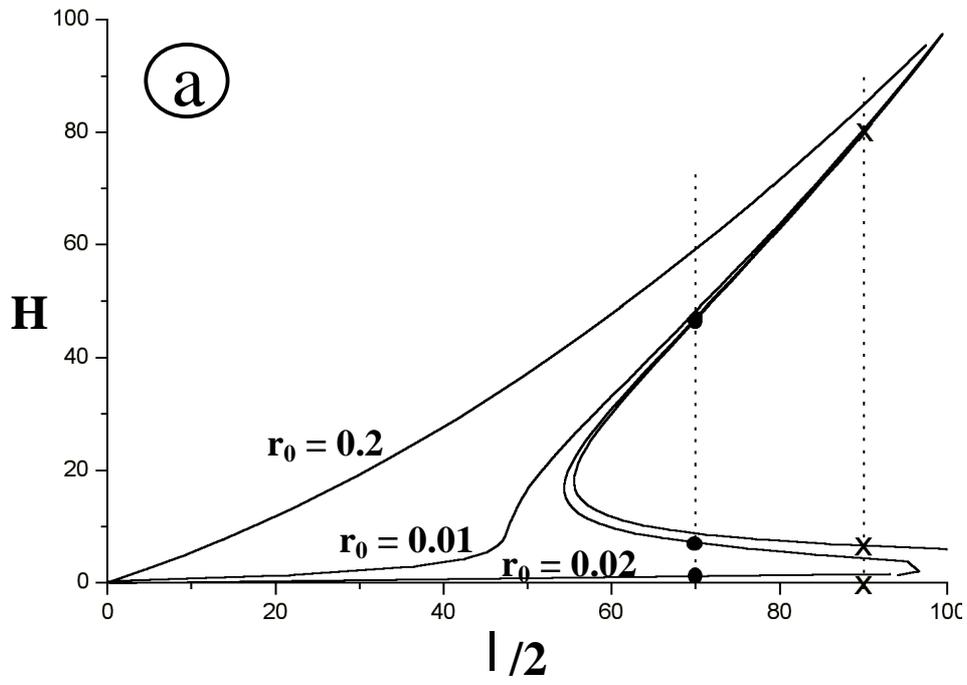

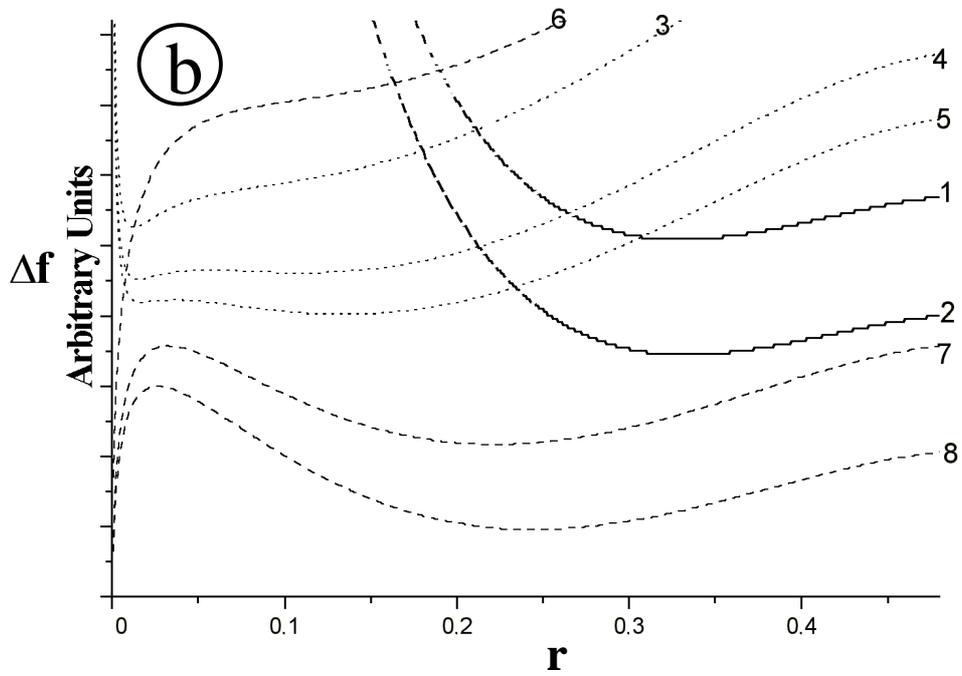

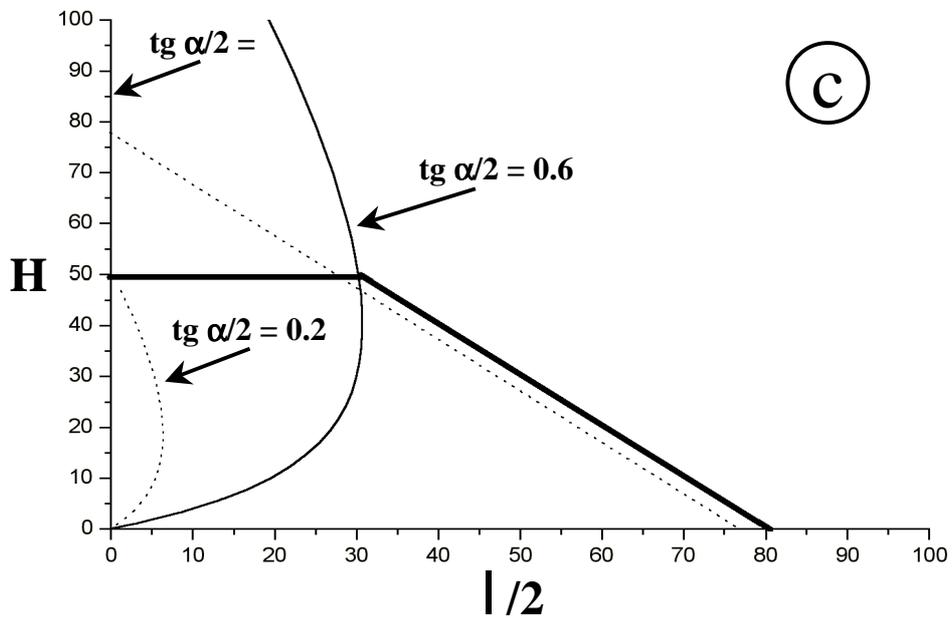

Figure 8 :

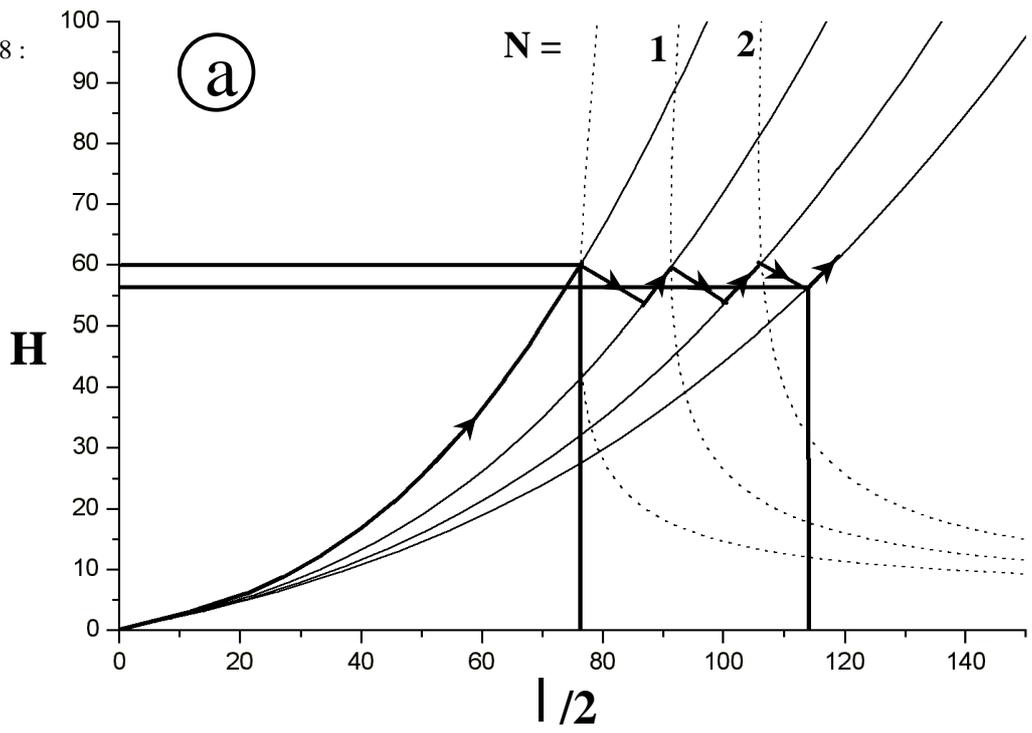

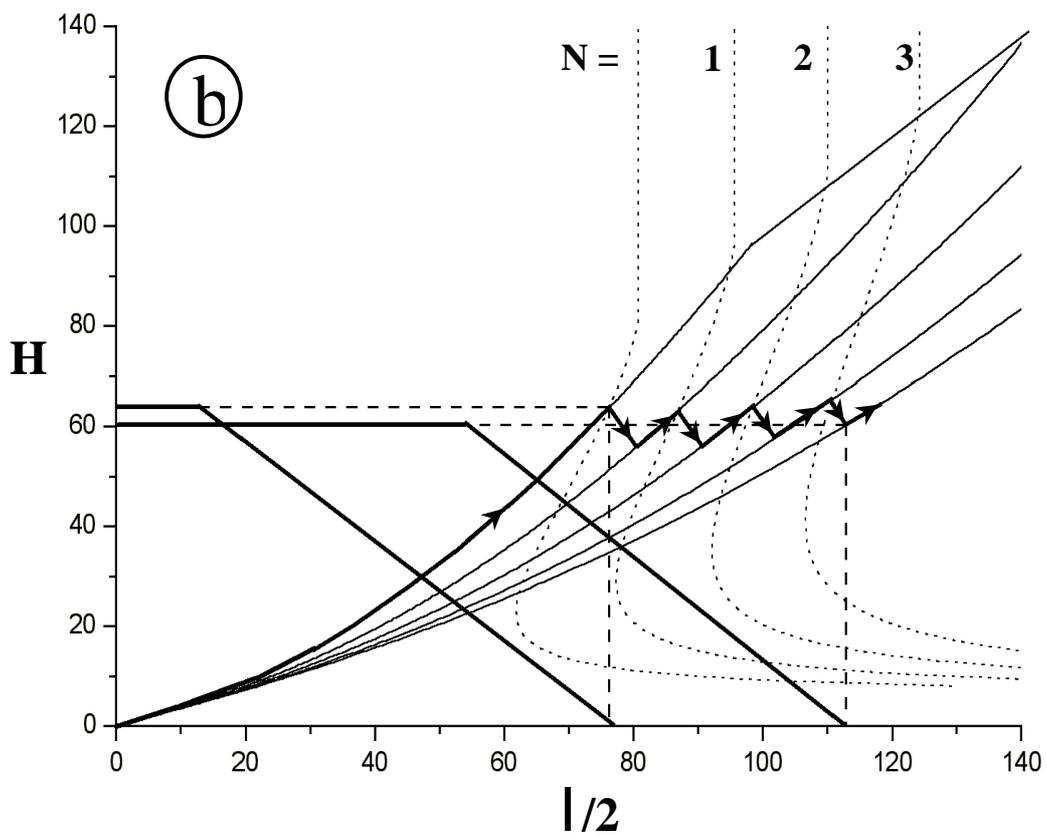

Figure 9 :